\documentclass[preprint]{aastex}

\newcommand{\Fig}[1]{Figure~\ref{#1}}
\newcommand{\Figo}[1]{Fig.~\ref{#1}}
\newcommand{\Figs}[2]{Figures~\ref{#1} and \ref{#2}}
\newcommand{\Figabc}[2]{Figure~\ref{#1}\textit{#2}}
\newcommand{\Figoabc}[2]{Fig.~\ref{#1}\textit{#2}}
\newcommand{\Eq}[1]{Equation~(\ref{#1})}
\newcommand{\eq}[1]{equation~(\ref{#1})}
\newcommand{\eqo}[1]{eq.~[\ref{#1}]}
\newcommand{\eqs}[2]{equations~(\ref{#1}) and (\ref{#2})}
\newcommand{\eqthree}[3]{equations~(\ref{#1}), (\ref{#2}), and (\ref{#3})}
\newcommand{\vct}[1]{\mbox{\boldmath{$#1$}}}
\newcommand{\mach}{\mathcal{M}}
\newcommand{\step}{\mathcal{H}}
\newcommand{\A}{\mathcal{A}}
\newcommand{\Amod}{\mathcal{\eta}}
\newcommand{\ext}{{\rm ext}}
\newcommand{\cs}{a_{\rm \infty}}
\newcommand{\tcr}{t_{\rm cross}}
\newcommand{\rs}{r_s}
\newcommand{\rA}{r_{\rm A}}
\newcommand{\rB}{r_{\rm B}}
\newcommand{\rBHL}{r_{\rm BHL}}
\newcommand{\rmin}{r_{\rm min}}
\newcommand{\ex}{\vct{\mathrm{e_x}}}
\newcommand{\ey}{\vct{\mathrm{e_y}}}
\newcommand{\ez}{\vct{\mathrm{e_z}}}
\newcommand{\eR}{\vct{\mathrm{e_R}}}
\newcommand{\Fplin}{F_{\rm lin}}
\newcommand{\aplin}{\alpha}
\newcommand\sles{\lower2pt\hbox{$\buildrel {\scriptstyle <}
   \over {\scriptstyle\sim}$}}

\shorttitle{NONLINEAR DYNAMICAL FRICTION}
\shortauthors{H.~KIM, \& W.-T.~KIM}

\begin{document}
\title{Nonlinear Dynamical Friction in a Gaseous Medium}
\author{Hyosun Kim and Woong-Tae Kim}
\affil{Department of Physics and Astronomy, FPRD, Seoul National University, Seoul 151-742, Korea}\email{hkim@astro.snu.ac.kr, wkim@astro.snu.ac.kr}

\begin{abstract}
Using high-resolution, two-dimensional hydrodynamic simulations,
we investigate nonlinear gravitational responses of gas to, and the
resulting drag force on, a very massive perturber $M_p$
moving at velocity $V_p$ through a uniform gaseous medium of adiabatic
sound speed $\cs$. We model the perturber as a Plummer potential with
softening radius $\rs$, and run various models with
differing $\A=GM_p/(\cs^2\rs)$ and
$\mach=V_p/\cs$ by imposing cylindrical symmetry with respect
to the line of perturber motion.
For supersonic cases, a massive perturber quickly develops nonlinear
flows that produce a detached bow shock and a vortex ring, which is
unlike in the linear cases where Mach cones are bounded by
low-amplitude Mach waves.
The flows behind the shock are initially non-steady, displaying
quasi-periodic, overstable oscillations of the vortex ring and the shock.
The vortex ring is eventually shed downstream
and the flows evolve toward
a quasi-steady state where the density wake near the perturber
is in near hydrostatic equilibrium.
We find that the detached shock distance $\delta$ and the nonlinear drag force
$F$ depend solely on
$\Amod=\A/(\mach^2-1)$ such that $\delta/\rs=\eta$ and
$F/\Fplin=(\Amod/2)^{-0.45}$ for $\Amod>2$, where $\Fplin$ is
the linear drag force of Ostriker (1999). The reduction of
$F$ compared with $\Fplin$ is caused by front-back symmetry in the
nonlinear density wakes.
In subsonic cases, the flows without involving a shock
do not readily reach a steady state. Nevertheless, the subsonic density
wake near a perturber is close to being hydrostatic, resulting in
the drag force similar to the linear case.
Our results suggest that dynamical friction of a very massive object as in
a merger of black holes near a galaxy center will take considerably
longer than the linear prediction.

\end{abstract}


\section{INTRODUCTION}\label{sec:intro}

A massive object in orbital motion suffers from orbital decay due to
a negative torque caused by gravitational interaction with its own 
gravitationally induced wake created in the background medium. 
This process, commonly referred to as dynamical friction (DF), occurs 
in not only collisionless environments (e.g., \citealt{chandra}) 
but also collisional gaseous backgrounds (e.g., \citealt{ost99}).
The DF in a gaseous medium is of great importance in understanding the 
formation and evolution of planets, binary stars, 
supermassive black holes, etc.
For instance, gravitational interaction between a protoplanet and 
its environmental disk causes the former to migrate toward a central 
star, naturally explaining the presence of ``hot Jupiters'' found 
from radial velocity surveys \citep[][and references therein]{but06}.
The migration also helps a planet grow faster in mass by providing 
an expanded gas-feeding zone at an enhanced accretion rate, which
may overcome the failure of \textit{in situ} core-accretion scenario 
in building a giant planet within a typical disk lifetime 
\citep[see, e.g.,][]{ali05}. 
In the case of nuclear black holes in merging galaxies, they are
expected to first experience the DF to form a binary and then coalesce 
into a supermassive black hole by emitting gravitational waves.
Friction of nuclear black holes against the 
collisionless \emph{stellar} background appears to be inefficient
due to scattering and depletion of stars near the
black holes, which is known as the ``final-parsec problem'' 
\citep[see][and references therein]{mil03}. However, 
recent numerical $N$-body/SPH simulations show that the gravitational drag 
from the \emph{gaseous} background is sufficient to form a black-hole 
binary in a relatively short time ($\sim1-10$ Myrs) 
\citep[e.g.,][]{esc04,esc05,dot06,dot07,may07,cua09}.

Thanks to a seminal paper of \citet{ost99}, DF in a gaseous medium is well 
understood as long as density wakes have small amplitudes. 
Earlier theoretical work by \citet{dok64}, \citet{rud71}, 
and \citet{rep80} considered density wakes in a steady state
and found that the drag force vanishes for a subsonic perturber, while 
it becomes remarkably similar to the collisionless counterpart 
for supersonic cases.
Using a time-dependent linear perturbation theory, on the other hand,
\citet{ost99} found 
that the gravitational drag force on a point-mass perturber
with mass $M_p$ moving at velocity $V_p$ on a straight-line
trajectory through a uniform gaseous medium with density $\rho_\infty$ and
sound speed $\cs$ is given by
\begin{equation}\label{eq:f_pt_linear}
  \Fplin = \frac{4\pi\rho_\infty (GM_p)^2}{V_p^2}
  \left\{\begin{array}{ll}
     \frac{1}{2}\ln(\frac{1+\mach}{1-\mach})-\mach, & \mach<1,\\
     \frac{1}{2}\ln(1-\frac{1}{\mach^2})+\ln(\frac{V_p t}{\rmin}), & \mach>1,
  \end{array}\right.
\end{equation}
where $\mach\equiv V_p/\cs$ is the Mach number,
$t$ is the time elapsed after the introduction of the perturber, and
$\rmin$ is the minimum radius introduced to avoid the singularity in 
the force evaluation. \Eq{eq:f_pt_linear} shows that the gaseous DF force 
becomes identical to the \citeauthor{chandra} (\citeyear{chandra}) formula
for the collisionless drag for $\mach\gg1$, and is, albeit small, 
non-zero even for subsonic perturbers because the time dependency 
breaks the symmetry in the density wakes (see also \citealt{just90}). 
\Eq{eq:f_pt_linear} has been applied to various astrophysical 
situations including 
orbital decay of compact objects in accretion disks 
(e.g., \citealt{nar00,kar01})
and heating of an intracluster medium by
supersonically moving galaxies in clusters
(e.g., \citealt{elz04,fal05,kim05,kim07,con08}).

While the result of \citet{ost99} is valid in a strict sense only 
for a linear-trajectory perturber in a uniform medium, it has proven
to be applicable to more general cases.
For example, \citet{KK} considered a 
circular-orbit perturber with orbital radius $r_p$ in a uniform gaseous 
medium and found that \eq{eq:f_pt_linear} is a reasonable approximation 
to the gaseous drag force on it, provided $V_pt=2r_p$. 
\citet{san01} numerically found that the orbital decay of a Plummer sphere
with a softening radius $\rs$ in a radially-stratified medium 
is consistent with 
the prediction of \eq{eq:f_pt_linear}, if $V_pt/\rmin = (0.35r_p/\rs)^{2.34}$.
Also, \citet{bar07} showed that \eq{eq:f_pt_linear} remains valid 
even for a perturber with relativistic speed if the relativistic 
correction factors are included. While \citet{dot06} found 
that the orbital decay of black hole binaries took longer than 
the prediction of \eq{eq:f_pt_linear} for a single perturber,
the discrepancy between the numerical and analytical results 
can be reconciled, at least partly, by taking allowance for fact that 
an object in a binary experiences not only a negative torque due to its
own wake but also a positive torque from the companion wake. 
For an equal-mass binary, \citet{KKS} found that the positive torque 
is on average about 40\% of the negative torque.

Since the results of \citeauthor{ost99} are based on the assumption that 
density wakes remain in the linear regime, the validity of \eq{eq:f_pt_linear}
for very massive perturbers has yet to be seen. 
The strength of gravitational perturbations due to a body with mass $M_p$
can be measured by the dimensionless parameter
\begin{equation}\label{eq:A} 
  \A=\frac{GM_p}{\cs^2\rs}, 
\end{equation}
which roughly corresponds to the perturbed density at a distance $\rs$ 
from the perturber relative to the background density 
(e.g., \citealt{just90,ost99}), and is equal to the 
Bondi radius $\rB=GM_p/\cs^2$ relative to $\rs$.
For systems with $\A\gg1$, the density wakes are clearly in the nonlinear 
regime and the linear perturbation analyses are likely to fail.
Identifying $\rs$ with the gravitational softening radius of a perturber
(or, equivalently, its size), $\A$ is in the ranges of 
$\sim0.1-1$ for galaxies embedded in typical intracluster media,
$\sim10-100$ for protoplanets in protostellar disks, and
$\sim10^6-10^8$ for supermassive black holes near galaxy centers,
suggesting that the wakes of massive compact objects can readily 
be nonlinear.
Indeed, \citet{esc05} reported that the orbital 
decay time of supermassive black hole binaries with $\A\sim10-100$ 
depends on $M_p$ much less sensitively than the results of the linear theory, 
which may be caused primarily by the nonlinear effects.

In this paper, we investigate nonlinear DF of a very massive perturber 
in a gaseous medium using numerical hydrodynamic simulations. 
In order to isolate the effects of the perturber mass and its 
velocity on the DF force, 
we consider a perturber following a straight-line trajectory in a uniform
gaseous medium, similarly to in \citet{ost99}. 
We model the perturber as a Plummer sphere that does not possess
any solid surface and merely provides gravitational potential perturbations
to the background medium that would otherwise remain static and uniform;
to make contact with the results of the linear theory, 
we ignore the accretion of gas onto the perturber in the current work. 
Our primary objectives are to find the changes in distributions of density
wakes with $\A$ and $\mach$, and to quantify the resulting 
gravitational drag forces in comparison with the linear cases.

Nonlinear responses of a background to a massive perturber moving at 
a supersonic speed have been extensively studied in the context of 
the Bondi-Hoyle-Lyttleton (BHL) accretion (\citealt{hoy39,bon44,bon52}; 
see also review of \citealt{edg04} and references therein).
Unlike our models where mass accretion to a perturber is prohibited,
however, the BHL accretion problem considered a perturber containing 
a defined surface through which gas is either accreted or reflected. 
It was \citet{hun71,hun79} who first solved for BHL accretion flows 
numerically, finding that the collisional nature of gas supports 
a bow shock in front of a supersonic perturber. Later studies found 
that the BHL accretion flows exhibit unstable behaviors such as 
flip-flop motions of the accretion shocks and vortex shedding when 
the condition of axisymmetry is relaxed 
(e.g.,\citealt{mat87,fry88,taa88,mat89,ruf94,fog97,fog99,fog05}).
While these numerical works on the BHL accretion explored temporal
evolution and distribution of density wakes as well as nonlinear 
features in some great detail, they mainly concentrated on the gravitational
focusing and resulting accretion rate of gas onto the perturber. 
Although some authors 
(e.g., \citealt{shi85,sha93,ruf96}) presented values for aerodynamic
and gravitational drag forces, only a limited range of $\A$ was 
covered. In this work, we run a number of numerical simulations 
by varying $\A$ and $\mach$ systematically
in order to quantify the dependences of the gravitational 
drag force on these parameters. A brief comparison between our results
with those from the BHL accretion studies will be presented.

This paper is organized as follow:
In \S\ref{sec:code}, we describe numerical methods we employ for 
nonlinear simulations.
In \S\ref{sec:linear}, as a code test we revisit the cases with 
a spatially-extended, linear perturber with $\A=0.01$. 
We compare the resulting distributions of density and velocity wakes with 
those of analytical results and provide a way to handle the effect 
on the DF force of a softening radius which is necessary for numerical 
simulations.
Evolution and quasi-steady distributions of fully nonlinear density wakes
and the associated drag forces are presented in \S\ref{sec:nonl}. 
Finally, in \S\ref{sec:diss} we summarize our findings and discuss 
their astrophysical implications.

\section{NUMERICAL METHOD} \label{sec:code}

In this paper, we consider an unmagnetized, inviscid, non-radiating, 
non-self-gravitating gaseous medium and study its gravitational responses
to a massive perturber moving along a straight-line trajectory using
numerical simulations. 
The background gaseous medium is initially static and uniform with 
density $\rho_\infty$ 
and adiabatic sound speed $\cs$. We adopt an adiabatic equation of state 
with an index $\gamma=5/3$ throughout simulations. 
The simulations are carried out on a
two-dimensional $(R, z)$ plane in cylindrical symmetry,
where $R$ and $z$ denote the distances from and along the axis of symmetry,
respectively.
We assume that the perturber exerts only gravity to the surrounding medium
and does not possess any surface, so that neither accretion nor reflection 
of the gas is allowed. 
The perturber is modeled as a Plummer sphere with mass $M_p$ that is moving 
at a constant speed $V_p$ along the $R=0$ axis toward 
the positive $z$-direction, with the gravitational potential 
\begin{equation}\label{eq:Plummer} 
\Phi_P(R,z,t)=-\frac{GM_p}{(R^2+\rs^2+[z-V_pt]^2)^{1/2}}, \;\;\;{\rm for}
\;t\geq0,
\end{equation}
where $\rs$ is the softening radius.

We take $\rs$, $\cs$, and $\tcr=\rs/\cs$ as the units of length, velocity,
and time, respectively, in our simulations. 
Then, our models are completely parametrized by $\A$ and $\mach$. 
We run a total of 58 models with $\A$ varying from 0.01 to 600
and $\mach$ in the range of 0.5 to 4.0. 
We solve the basic equations of ideal hydrodynamics
using FLASH3 \citep{fry00}, an Eulerian hydrodynamics code 
that implements a direct Piecewise-Parabolic Method 
solver of \citet{col84} for high-order spatial reconstruction. 
Although the FLASH3 code is capable of both uniform grid and adaptive mesh 
refinement calculations,
we adopt the uniform grid method since the accurate evaluation of the
drag force requires the whole computational domain to be well resolved.
As we will show below, we find that it is necessary to 
have at least 5 zones per $\rs$ to obtain converged results for the 
drag forces. Our largest grid models have 3,072$\times$12,288 zones
in $(R,z)$; 
we make sure that our computational domain is taken to be large enough 
to contain the whole density wake in a given model.
The simulations are typically carried out until $t/\tcr=600$
when most of the wakes are well resolved and reach a quasi-steady state.

\section{LINEAR CASES} \label{sec:linear}

Time-dependent linear perturbation theories for the DF drag force in 
a gaseous medium usually study the responses of gas to a low-mass, point-mass 
perturber corresponding to $\rs=0$ in the Plummer potential (e.g., 
\citealt{just90,ost99}), which requires to introduce the cut-off radius 
$\rmin$ in the linear force formula (\eqo{eq:f_pt_linear}).
In numerical simulations, on the other hand, one needs to assign
a non-zero value to the softening radius, which in turn makes it
unnecessary to use the cut-off radius in the force evaluation.
Since our goal is to compare the nonlinear drag force on a massive 
perturber with the linear prediction, we have to first find a proper
relationship between $\rs$ and $\rmin$ that makes the numerical and 
analytical results consistent with each other when $A\ll1$.
Motivated by this consideration, in this section we briefly present 
the results of 
numerical simulations for a low-mass perturber with $\A=0.01$, and 
compare the resulting distributions of density and velocity wakes
and the drag forces with those from the linear theories. This will also
allow us to check the accuracy of our numerical experiments.

Regarding the linear wakes with which numerical results will be compared, 
it is worth mentioning that there are several analytical methods
for finding solutions for the perturbed density and velocity fields.
\citet{just90} utilized Fourier transform for the space variables 
and Laplace transform for the time variable, finding expressions
both for the density and velocity wakes. Instead, \citet{ost99} used a
retarded Green's function technique
and found an expression only for the density wake that is identical to
the result of \citet{just90}. We found that while the 
analytical formula for the density wake agrees well with our numerical
results for a low-mass perturber, the simulated velocity field differs
from the expression given by \citet{just90}.
In Appendix \ref{sec:FT}, we revisit the time-dependent linear theory
using Fourier transforms both for the space and time variables.
As we will show below, our expressions (\ref{eq:vz}) and (\ref{eq:vR}) for
the perturbed velocities are in good agreement with the numerical results,
confirming that equation~(47) of \citet{just90} contains a typographical
mistake.

\begin{figure}
  \epsscale{1}
  \plotone{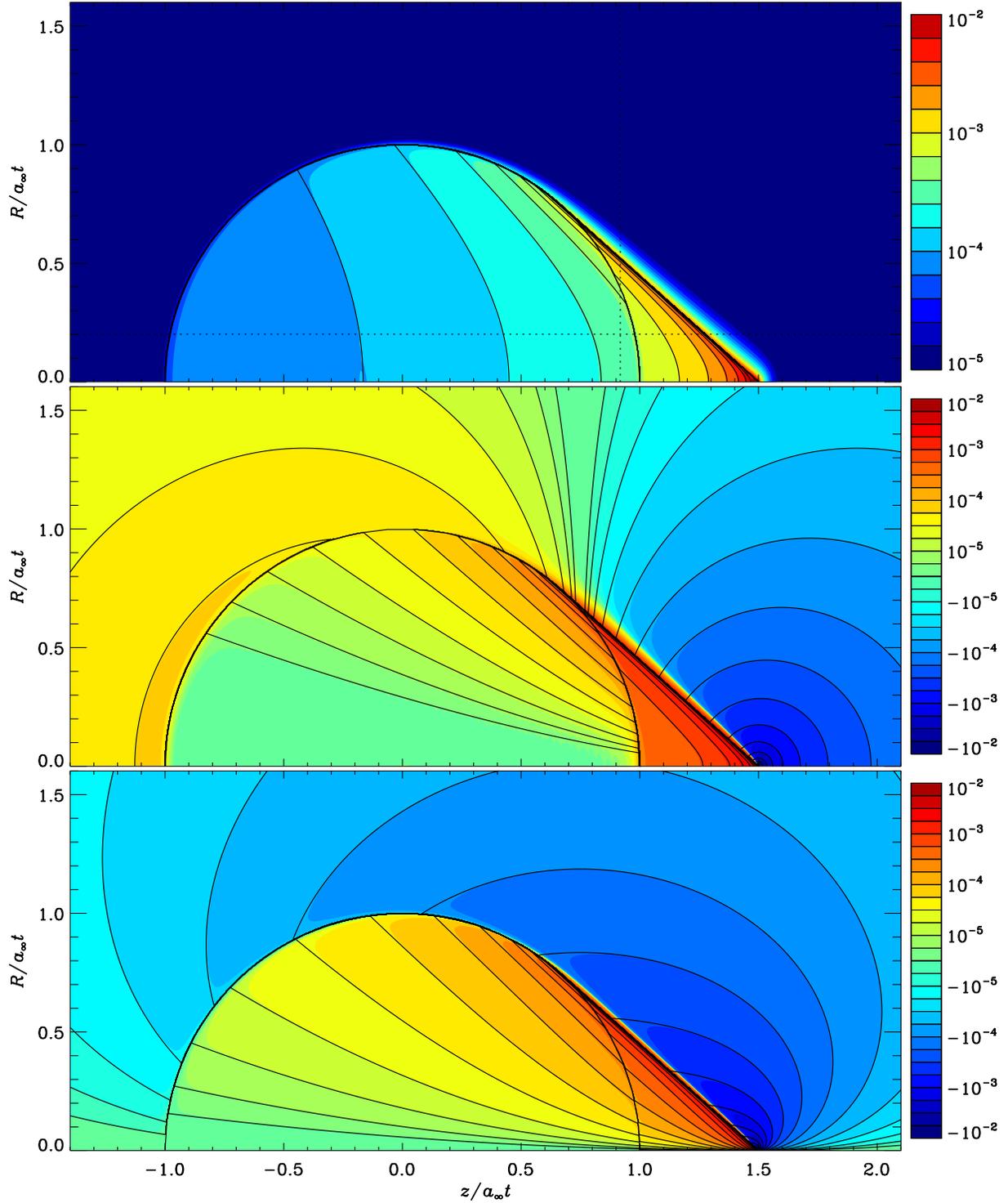}
  \caption{\label{fig:linear}
  Color-scale distributions of the perturbed density $\alpha$ (\textit{top}), 
  parallel velocity $v_z$ (\textit{middle}), and perpendicular velocity 
  $v_R$ (\textit{bottom}) to the line of motion of an extended perturber 
  with $\A=0.01$ and $\mach=1.5$. 
  Colorbars label $\alpha$, $v_z/\cs$, and $v_R/\cs$ from
  top to bottom.
  The black contours represent the results of the time-dependent linear
  perturbation theory for the corresponding point-mass perturber.
  The perturber initially located at $(z,R)=(0,0)$ has moved to 
  $(\mach\cs t, 0)$ at time $t$ along the $z$-axis.}
\end{figure}

\Fig{fig:linear} shows as color-scale images the snapshots on the
$z$--$R$ plane 
of the perturbed density $\alpha=\rho/\rho_\infty-1$ (\emph{top}),
the parallel velocity $v_z$ (\emph{middle}), and
the perpendicular velocity $v_R$ (\emph{bottom}) to the line of motion 
for a model with $\A=0.01$ and $\mach=1.5$. 
Note that the coordinates are normalized by $\cs t$.
The perturber initially
introduced at $(z,R)=(0,0)$ has moved to $(\mach\cs t, 0)$ at time $t$.
The characteristic features of a supersonic wake consisting
of a sonic sphere with radius $\cs t$ 
centered at the initial perturber location and a Mach cone bounded by Mach 
waves located at $R=-(\mach^2-1)^{-1/2}(z-\mach\cs t)$ 
for $z>\cs t/\mach$ are apparent in the top panel. Also plotted 
as black solid contours are the results of the linear perturbation
theory (eqs.\ [\ref{eq:alpha}], [\ref{eq:vz}], and [\ref{eq:vR}]),
which are overall in good agreement with the numerical results.
A careful inspection of \Fig{fig:linear}, however, reveals that
the numerical results deviate slightly from the analytical ones especially
near the sonic radius and the Mach waves.
This is due to the fact that the perturber in our numerical models 
is modeled as an extended Plummer sphere rather than a point mass.

\begin{figure}
  \epsscale{1}
  \plotone{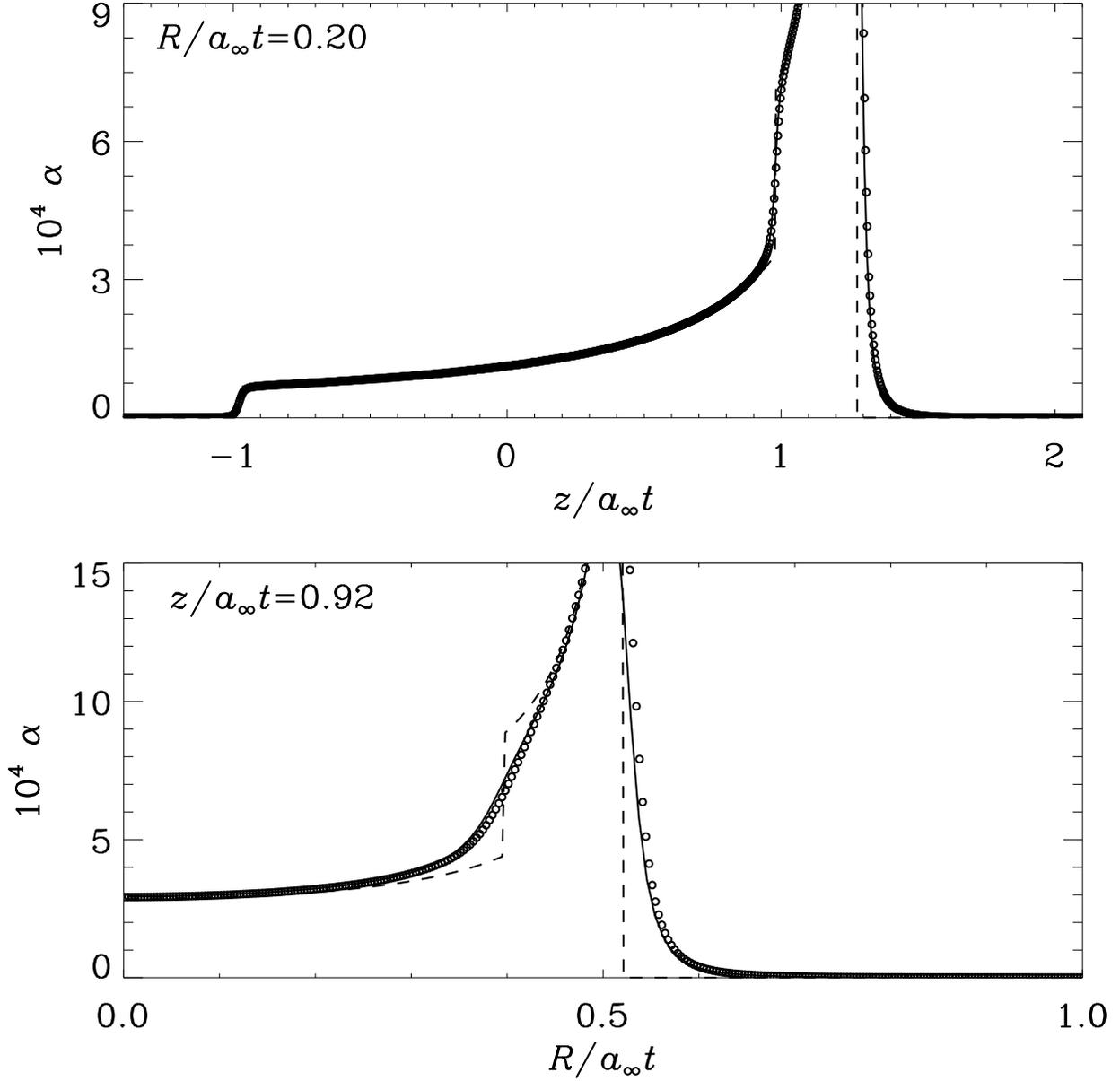}
  \caption{\label{fig:linprof}
  Distributions of the perturbed density along the cuts at $R/\cs t=0.20$ 
  (\textit{top}) and $z/\cs t=0.92$ (\textit{bottom}) indicated as dotted lines 
  in \Figo{fig:linear} of a model with $\A=0.01$ and $\mach=1.5$. Open 
  circles representing the simulation results deviate considerably from 
  the results $\alpha$ (\textit{dashed line}) of the linear perturbation 
  theory for a point mass, but are in excellent agreement with 
  $\alpha_{\rm ext}$ (\textit{solid line}) obtained from the convolution 
  of $\aplin$ with the extended mass distribution.} 
\end{figure}

One can semi-analytically construct the density wake $\alpha_{\rm ext}$ 
of an extended perturber by convolving the density wake $\aplin$ 
due to the corresponding point mass with the extended 
mass distribution $\rho_{\rm ext}$ of the perturber 
(e.g., \citealt{just90,fur02}). 
For a Plummer sphere we use, the convolution theorem gives
\begin{equation}\label{eq:cnv}
  \alpha_{\rm ext}(\vct{x},t)=
  \frac{1}{M_p} \int
    \alpha(\vct{x}-\vct{x^\prime},t)\rho_{\rm ext}(\vct{x^\prime},t)
    \;d^3\!\vct{x^\prime},
\end{equation}
where $\rho_{\rm ext} 
(\vct{x},t)=3M_p\rs^2(R^2 + \rs^2 +(z-V_pt)^2)^{-5/2}/(4\pi)$.
\Fig{fig:linprof} plots as solid lines the profiles of 
$\alpha_{\rm ext}$ for $\A=0.01$ and $\mach=1.5$ along the cuts 
at $R/\cs t=0.20$ and $z/\cs t=0.92$ marked as dotted lines in the top 
panel of \Fig{fig:linear}, which are in excellent agreement with
the simulation outcomes (\emph{open circles}).
Compared with the point-mass results (\emph{dashed lines}),
the extended mass distribution tends to smear out the discontinuities 
at the boundary of the sonic sphere and the Mach cone. This makes sense
since a perturbed density at one location is a superposition of
sonic perturbations with various strengths launched by 
all the mass elements comprising the extended body.

\begin{figure}
  \epsscale{1}
  \plotone{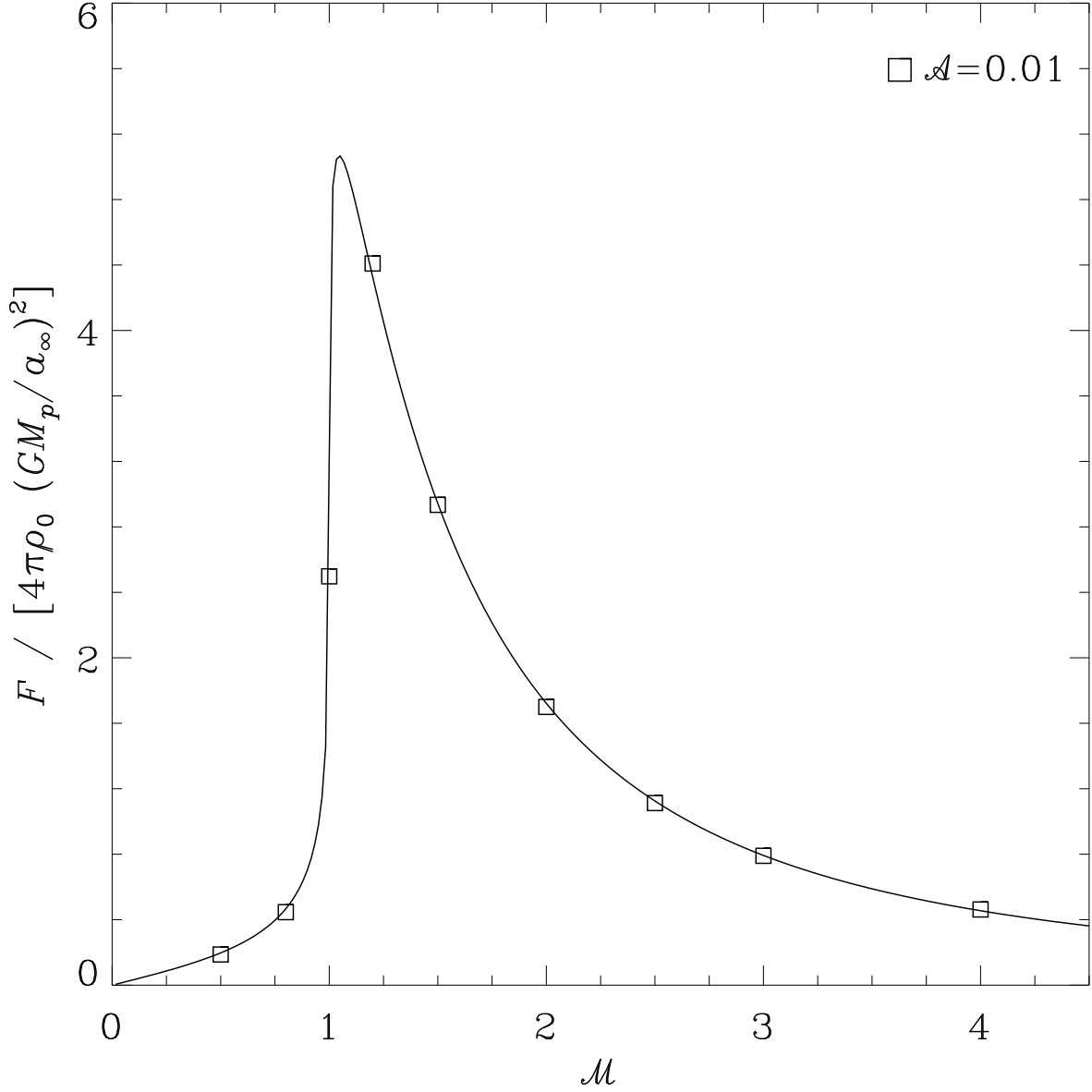}
  \caption{\label{fig:fMlin}
  Dimensionless DF force on a low-mass perturber with $\A=0.01$ 
  against the Mach number $\mach$ at $t/\tcr=300$. 
  The solid line is the best fit of the linear force formula 
  (\eqo{eq:f_pt_linear}) with $\rmin=0.35\mach^{0.6}\rs$ to
  our simulation results (\textit{open squares}).}
\end{figure}

Allowing for the extended mass distribution, the gravitational drag force 
exerted on the perturber in the negative $z$-direction can be obtained by 
directly evaluating the integral
\begin{equation}\label{eq:f_integ}
  F(t)=\int\!\!\!\!\int
  \frac{G\rho_{\rm ext}(\vct{x^\prime},t)\rho(\vct{x},t) (z-z^\prime)}%
       {|\vct{x}-\vct{x^\prime}|^3}\ d^3\vct{x}\ d^3\vct{x^\prime},
\end{equation}
where $\rho(\vct{x},t)$ is the wake distribution.
Although \eq{eq:f_integ} requires the $\vct{x^\prime}$-integration 
to be performed over the entire Plummer sphere,
we empirically found that the drag force on 
the region with distance from the perturber larger than $10\rs$ 
has a negligible contribution to the total. 
Thus, in practice, 
we limit the integration to the region within $10\rs$ that contains 
about 98.5\% of the total perturber mass.
\Fig{fig:fMlin} plots as open squares the numerical drag forces 
on a low-mass perturber with $\A=0.01$ but differing $\mach$. 
The solid line corresponds to \eq{eq:f_pt_linear} with 
\begin{equation}\label{eq:rmin}
r_\mathrm{min}=0.35\mach^{0.6}\rs,
\end{equation}
which gives the best fit to the supersonic results of our adiabatic 
simulations. 
\Eq{eq:rmin} is our prescription for the cut-off radius when we compare
the numerical drag forces with the analytic results. 
Note that \citet{san99} suggested $\rmin=2.25\rs$
based on their isothermal simulations.  
The difference between the two prescriptions may be due in part 
to using different equations of state and in part to low resolution (1 cell 
per $\rs$) in \citet{san99}.

\section{NONLINEAR CASES} \label{sec:nonl}
\subsection{Wake Evolution} \label{sec:general}
\subsubsection{Supersonic Cases}
\begin{figure*}
  \epsscale{1}
  \plotone{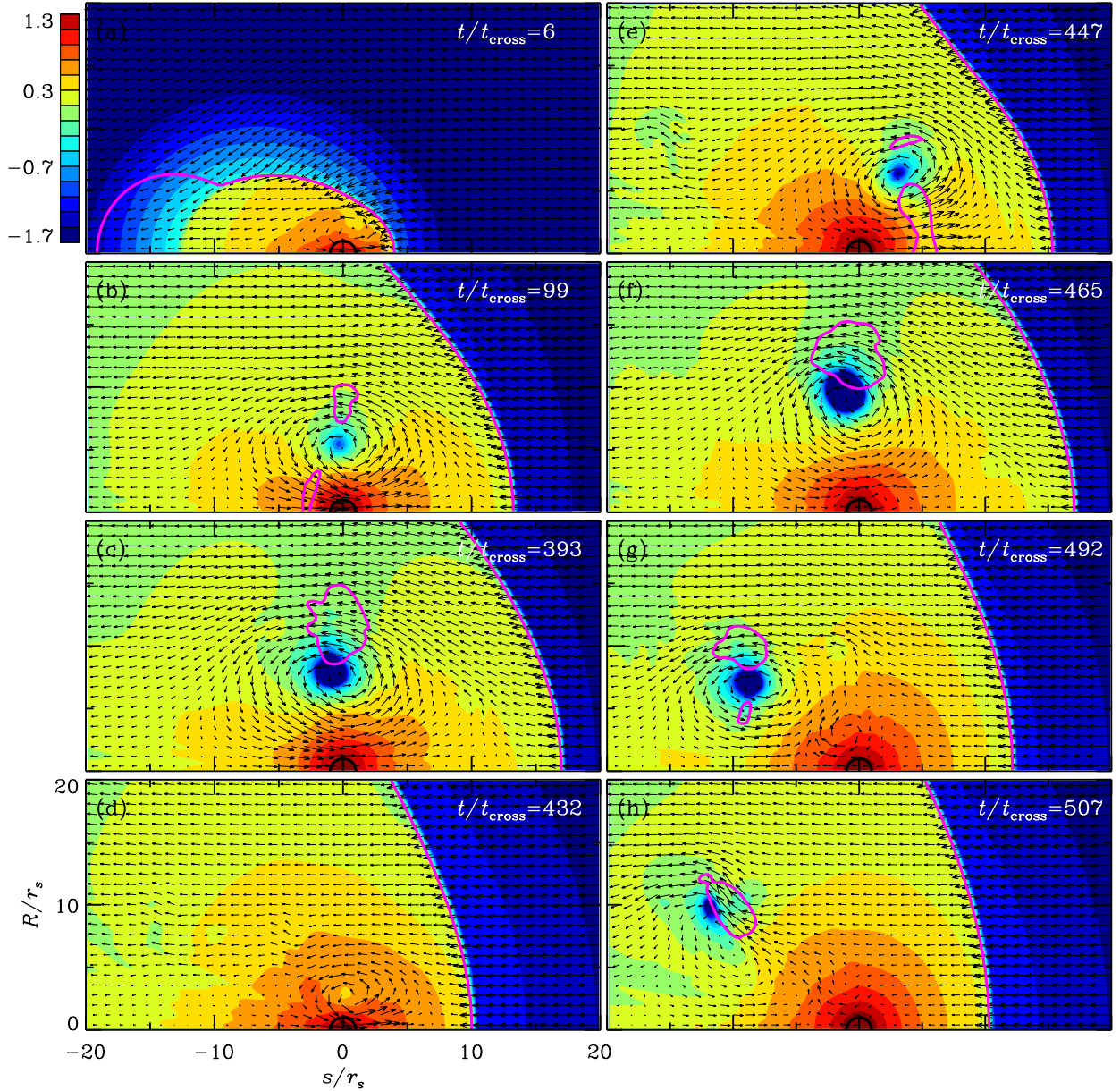}
  \caption{\label{fig:evol}
  Temporal evolution of the density structure (logarithmic color-scale) and
  velocity field (arrows) generated by an extended 
  perturber with $\A=20$ and $\mach=1.5$ in the frame where the 
  perturber located at $(s=z-V_p t, R)=(0,0)$ is stationary.
  A black semicircle marks the softening radius 
  of the perturber, while red curves draw the sonic lines where 
  the gas speed is equal to the local sound speed.
  Colorbar labels $\log (\rho/\rho_\infty-1)$.}
\end{figure*}

We begin by describing the temporal evolution of 
our fiducial model with $\A=20$ and $\mach=1.5$; 
the evolution of other supersonic models are qualitatively similar. 
\Fig{fig:evol} illustrates the density and velocity structures 
of this model in a comoving frame with the perturber located at 
$(s\equiv z-V_pt, R)=(0,0)$. Only the region with
$ |s|/\rs \leq 20$ and $0 \leq R/\rs \leq 20$ is shown.
Unlike the linear cases with $\A\ll1$ where the density wakes
have too small amplitudes to launch shock waves, 
the perturber with $\A=20$ emits strong perturbations that quickly 
develop into a bow shock. 
The upstream gas moving toward the perturber along the symmetry axis 
is first accelerated by the gravity of the perturber, is shocked to 
subsonic speed, and then piles up near the center of the perturber.
This creates a steep pressure gradient in between the perturber and 
the shock, tending to push the shock away from the perturber.
On the other hand, the gas flowing above (not far away from) 
the symmetry axis is deflected toward the perturber even before 
entering the shock and decreases its speed after the shock. This gas
thus has a longer time to be exposed to the gravity of the perturber 
as it moves toward the symmetry axis. 
The gravitational potential is so deep that the material arriving at 
the rear side of the perturber can be pulled back toward the perturber,
creating a stagnation point just as in the BHL accretion flows 
(e.g., \citealt{mat87,fry87}).
This produces a strong counterstream that moves into the upstream direction
along the symmetry axis, as well as a primary vortex in the $s$--$R$ plane 
(\Figoabc{fig:evol}{a})\footnote{More precisely, the vortex in the $s$--$R$ 
plane is a vortex ring in three dimensions, and the counterstream is a part 
of the vortex ring near the symmetry axis.}. The counterstream combined with 
the pressure gradient in the front side pushes the shock front away from 
the perturber. 

\begin{figure}
  \epsscale{1}
  \plotone{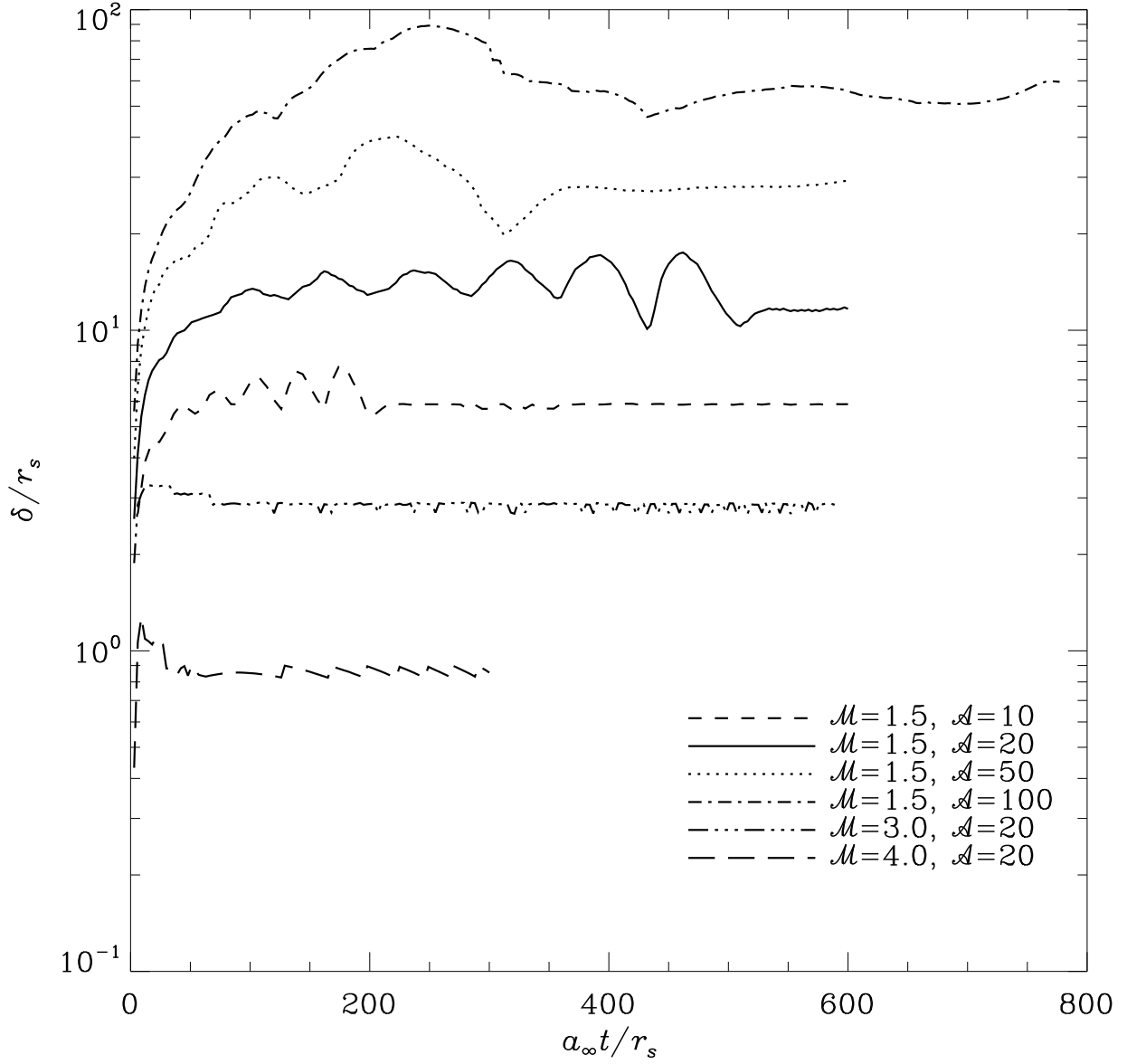}
  \caption{\label{fig:dtime}
  Time evolution of the detached shock distance $\delta$ measured
  along the symmetry axis from the perturber center for various models. 
  For each model with $\A\le50$, 
  $\delta$ initially increases rapidly with time, experiences 
  quasi-periodic oscillations, and then saturates to a constant value. 
  The time to reach a quasi-steady state and the value of $\delta$ at 
  saturation become smaller as $\A$ decreases or $\mach$ increases; 
  the model with $\A=100$ and $\mach=1.5$ does not attain a 
  quasi-steady state until the end of run.}
\end{figure}

\begin{figure}
  \epsscale{1}
  \plotone{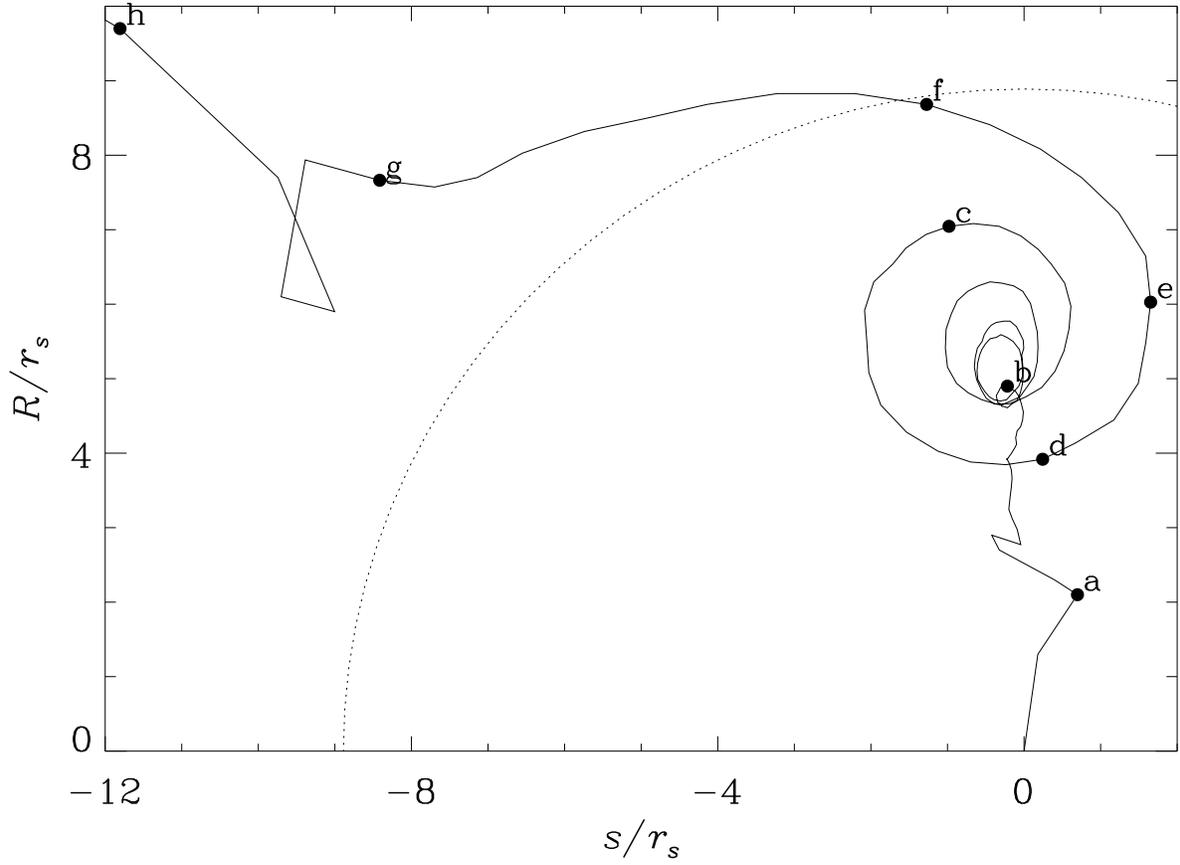}
  \caption{\label{fig:vort}
  Trajectory of the center of the primary vortex on the $s$--$R$ plane 
  for a model with $A=20$ and $\mach=1.5$. 
  Filled circles marked by alphabets a--h 
  indicate the vortex locations at time epochs corresponding 
  to the snapshots shown in \Figo{fig:evol}. The dotted curve draws
  the region where the distance from the perturber equals 
  a half of accretion radius $\rA\equiv 2GM_p/V_p^2$.}
\end{figure}

The advance of the shock front in the upstream direction allows 
more time for the shocked gas to be affected by the gravity, 
strengthening the counterstream (as well as the primary vortex) and 
thus increasing the detached distance $\delta$ of 
the shock measured along the symmetry axis. 
At the same time, the center of the vortex with a lower density
than the surrounding buoyantly rises toward the high-$R$ regions.
This decreases the ram pressure of the counterstream exerted on the shock 
front and the shock advance slows down. 
\Fig{fig:dtime} plots the time evolution of $\delta$ for some 
selected models, while \Fig{fig:vort} traces the trajectory of the center
of the primary vortex in our standard model ($\A=20$ and $\mach=1.5$) 
on the $s$--$R$ plane. 
For our standard model, $\delta$ keeps increasing to $13\rs$ at $t/\tcr=100$. 
At this time, the vortex is located near at $(s,R)\simeq(-0.2\rs,5\rs)$ 
(Figs.\ \ref{fig:evol}{b} and \ref{fig:vort}).
The shock front soon overshoots a potential equilibrium position where 
the thermal pressure of the postshock gas supports it against the
perturber gravity, and begins oscillating around the equilibrium position. 

The shock oscillation changes the velocity field in the postshock
region, causing the vortex to move in the counterclockwise direction
around its mean position $(-0.5\rs, 5\rs)$ following the background flow.
Kelvin's circulation theorem implies that the vortex becomes
stronger as it moves toward the symmetry axis, amplifying the strength 
of the counterstream near the perturber.
The velocity of the counterstream becomes largest when the vortex arrives 
closest to the perturber on its oscillation path (\Figoabc{fig:evol}{d}). 
The shock front that was moving toward the perturber is pushed by the
strong counterstream, and reverses its motion.
The vortex rises to high $R$ as the shock moves away from the perturber,
weakening the counterstream, and the 
oscillation cycle repeats quasi-periodically. 

\Figs{fig:dtime}{fig:vort} show that the amplitudes of the detached shock 
oscillations as well as the vortex movements grow secularly with time 
as the shock oscillation continues. This can be understood as follows.
When the shock front is displaced from the equilibrium position away
from the perturber (e.g., \Figoabc{fig:evol}{c}), 
the shocked subsonic gas can acquire an
extended time to be gravitationally influenced, similarly to the
early situation when the shock advances away from the perturber. 
With the stagnation point moving away from the perturber, more mass
and momentum are added to the counterstream, amplifying 
the vortex oscillation. In addition,
when the strong counterstream collides directly with the shocked subsonic 
flow near the symmetry axis, the material at the interface 
is injected upward in the lateral direction and then carried in the
negative $s$-direction by the background flow (see, e.g., the region 
at $s/\rs\sim12$ and $R/\rs\sim0-3$ in \Figoabc{fig:evol}{e}). 
This produces small vortices near the interface that move downstream 
and merge with the primary vortex, again strengthening the latter.
Consequently, the primary vortex is able to move
closer to the symmetry axis and amplify the counterstream in
the next cycle, making the shock oscillation overstable.

As the center of the primary vortex moves away from the symmetry axis 
during its last phase of the overstable oscillation ($t/\tcr\sim465-492$),
it becomes less gravitationally bound and can thus be more easily 
influenced by the background flow. For all the \emph{supersonic} models we run
in this paper, we find that whenever the vortex rises above a half
of the accretion radius, also known as the Hoyle-Lyttleton radius,
defined by $\rA\equiv 2GM_p/V_p^2$, it is swept away by the 
background flow in the negative $s$-direction (\Figo{fig:vort}). 
When the primary vortex is located outside $0.5\rA$,
the counterstream associated with the vortex is weak and 
occurs in the far rear side of the perturber, unable to 
pass through the center of the perturber.
The counterstream is instead resisted by the flow moving downstream
right across the perturber and pushed up in the lateral direction
(\Figoabc{fig:evol}{g}). 
This develops another vortex with an opposite sense of rotation 
to the primary one, capable of pushing the latter away from 
the perturber (\Figoabc{fig:evol}{h}).
With the primary vortex carried away downstream, the flow near the 
perturber reaches a quasi-steady state in which shocked subsonic gas
moves almost parallel to the symmetry axis. Since the associated
kinetic energy is much smaller than the thermal and gravitational energies
of the gas, the density distribution around the perturber becomes nearly
hydrostatic, as will be shown below.

\subsubsection{Subsonic Cases}

\Fig{fig:subs} displays density and velocity structures of 
a nonlinear subsonic model with $\A=20$ and $\mach=0.5$
in a comoving frame with the perturber. 
Snapshots at $t/\tcr=60$, 150, and 600 are shown.
Sound waves launched from the perturber at $t=0$ propagate radially
outward into the surrounding medium, forming a spherical 
causal region within which the medium is affected by the sonic 
perturbations. Since the perturber is spatially extended, however, 
the boundary of the casual region is not as sharp as in the case of a 
point mass, although the most dominant perturbations 
still come from the perturber center.
Unlike supersonic cases, this model always involves subsonic flows and 
never produces a shock. Nevertheless, 
the overall flow pattern and late-time density structure near the perturber
of this model is very similar to those in the postshock subsonic regions 
of supersonic models. First of all, 
the strong gravitational pull forms a counterstream and an
associated vortex ring near the symmetry axis (\Figoabc{fig:subs}{a}).
The counterstream moving in the upstream direction interacts with
the incident flow (\Figoabc{fig:subs}{b}).  
The gas at the interface is pushed up toward the high-$R$ regions
and then carried downstream, creating small vortices 
with low density (\Figoabc{fig:subs}{c}).  

The primary vortex slowly rises in the $R$-direction due to buoyancy, 
and merge with the small vortices. 
Since this model does not contain a shock that would confine the region 
of influence and since the causal region keeps expanding at a sonic speed, 
the flows in the high-$R$ regions are almost parallel to the symmetry axis. 
As a result, the primary vortex keeps rising as there is no momentum input 
in the background flow capable of pushing it back toward the 
symmetry axis. 
At the end of the run ($t/\tcr=600$), the primary vortex in this model 
arrives at $(s,R)=(3\rs, 28\rs)$, corresponding to $0.18\rA$.
It is uncertain whether the primary vortex will be carried downstream
when it goes beyond $r=0.5\rA$ in a manner similar to supersonic cases.
At any event, the density distribution close to the perturber is well 
described by the condition of hydrostatic equilibrium at late time 
(see \S\ref{sec:wake} below).

\subsection{Quasi-steady Density Wakes}\label{sec:wake}

\Fig{fig:compAdens} 
illustrates changes in the quasi-steady density wakes with varying $\A$ 
on the $s$--$R$ plane for supersonic models with fixed $\mach=1.5$ at
$t/\tcr=600$. 
The left panels show large-scale views of the wakes at
$ -1400 \leq s/\rs \leq 100$ and $0 \leq R/\rs \leq 600$,
while the region near the perturber 
with $ -60 \leq s/\rs \leq 20$ and $0 \leq R/\rs \leq 60$ 
is enlarged in the right panels.
In each panel, the perturber is located at $s=R=0$,
and the black line connecting the points $(s,R)/\rs=(0,0)$ and
$(-60,54)$ marks the boundary of the Mach cone characteristics of 
the linear density wake for $\mach=1.5$.
When $\A \ll1$, the sonic perturbations are too weak to produce a shock,
and the high-density ridge of the wake follows the Mach cone fairly well, 
although it is broadened due to the extended mass distribution of the
perturber. 
As $\A$ increases to unity, sonic perturbations even outside the Mach
cone attain substantial amplitudes enough to induce a bow shock that is 
attached to the center of the perturber (within the resolution limit).
Since the density wake is effectively shifted toward the perturber 
compared with the linear case and still located preferentially 
at the rear side of the perturber, the resulting drag force will 
be larger than the linear counterpart (see \S\ref{sec:force}).

As $\A$ increases further, the shocked material gathered around the perturber 
begins to build up a strong pressure barrier which the incident flow cannot 
easily penetrate. This naturally makes the shock detached. 
\Figabc{fig:compAdens}{b,c} shows that the postshock density 
distribution around the perturber with $\A=10$ or 20 is almost spherically 
symmetric, indicating that the kinetic energy of the gas there is much 
smaller than the thermal and gravitational potential energies. 
To check if this is indeed the case, we plot in \Fig{fig:HSE} the density 
profiles along the symmetry axis and the $R$-axis from the center of 
a massive perturber with $\A=20$; both 
supersonic and subsonic models with $\mach=1.5$ and $0.5$
at $t/\tcr=600$ are presented. 
Also shown as dotted line is the density distribution
under the assumption of hydrostatic equilibrium
\begin{equation}\label{eq:HSE}
\rho=\rho_0 
  \left\{1+\frac{(\gamma-1)\A\cs^2}{a_0^2}%
       \left[\frac{\rs}{(r^2+\rs^2)^{1/2}}-1\right]%
  \right\}^{1/(\gamma-1)},
\end{equation}
where $r=(s^2+R^2)^{1/2}$ denotes the distance from the perturber center
and $\rho_0$ and $a_0$ are the density and the adiabatic sound speed
at $r=0$, respectively.

For both supersonic and subsonic models, the density distributions 
at $r/\rs\la 10$ 
along the $z^+$-, $z^-$-, and $R$-cuts are virtually identical to
each other, and are in remarkable agreement with the predictions of 
\eq{eq:HSE} with $\rho_0/\rho_\infty=40$ and $a_0/\cs=3.8$ for 
$\mach=1.5$, and $\rho_0/\rho_\infty=52$ and $a_0/\cs=3.8$ for $\mach=0.5$. 
The sharp drop offs of the density in the supersonic model near $r/\rs=12$ 
and $32$ along the $z^+$- and $R$-directions, respectively, are 
of course due to the bow shock. 
In the subsonic model, the presence of the primary vortex 
makes the local density decreased at $R/\rs=28$. 
Note that the quantity $a^2/\rho^{\gamma-1}$ measures the specific entropy 
and thus is conserved in an adiabatic flow without involving a shock,
as is the case in the subsonic model. For the supersonic model, however, 
$a^2/\rho^{\gamma-1}$ is increased from unity to 1.34 because of a shock jump.
This corresponds to the shock Mach number of $2.3$, which is larger than 
$\mach=1.5$ because the flow is accelerated by the perturber even before 
experiencing the shock.
In the regions with $r/\rs \ga 10$, the density is overall larger along 
the $z^-$- than $z^+$-directions, providing non-vanishing drag forces.
Nonetheless, the presence of hydrostatic cores in the density wakes of
massive supersonic perturbers makes the drag forces smaller than 
the linear results. 

While the density wake \emph{near} a massive perturber with
$\A\geq1$ is quite different 
from the linear counterpart, the distant part of the wake is more or less
the same.
\Fig{fig:profiles} plots the exemplary profiles of the normalized
perturbed density, $\alpha/\A$, along the cuts with $R=0$ and $R/\rs=200$ 
for models shown in \Fig{fig:compAdens}.  Note that 
$\alpha/\A$ is nearly independent of $\A$ in regions far away
from the perturber (e.g., $s/\rs < -900$ region in \Figoabc{fig:profiles}{a}
and $s/\rs < -500$ region in \Figoabc{fig:profiles}{b});
in these regions, the gravitational potential 
perturbations are weaker by more than two orders of magnitudes than
at the perturber location and thus locally in the linear regime. 
Even with low amplitudes of local perturbations, however, the region with 
$ -900 < s/\rs < -100$ along the symmetry axis behind the perturber 
has the nonlinear density wake that deviates considerably from the linear case.  This is
because the gas flowing in this region was already affected by strong
gravitational potential in the upstream region, and has a diverging
velocity field that reduces the perturbed density.  Small fluctuations 
of nonlinear density wakes apparent in \Figs{fig:compAdens}{fig:profiles}
near the $R=0$ axis is thought of as arising from sonic perturbations 
induced by the primary vortex and its oscillations discussed 
in \S\ref{sec:general}.

\subsection{Detached Shock Distance} \label{sec:delta}

We have shown in the previous subsection that a sufficiently-massive
supersonic perturber generates a density wake that is characterized by a bow 
shock standing ahead of the perturber and a surrounding hydrostatic envelope.
\Fig{fig:dtime} shows that the quasi-steady value of the detached 
shock distance $\delta$ is larger for models with larger $\A$ or smaller 
$\mach (>1)$. To quantify the dependences of $\delta$ upon $\A$
and $\mach$, we introduce the nonlinearity parameter
\begin{equation}\label{eq:Amod}
  \Amod\equiv \frac{\A}{\mach^2-1},
\end{equation}
and plot in \Fig{fig:dEta} the normalized shock distance against $\Amod$.
The various symbols give the mean values of $\delta$ 
temporally averaged over $t/\tcr>50$ that ignores the initial 
wake-development phase. The standard deviations of $\delta$ are also 
indicated by errorbars. The numerical results are remarkably well described by
the two simple power laws: $\delta/\rs=2(\Amod/2)^{2.8}$ for 
$0.7\la\Amod\la2$ and $\delta/\rs=\Amod$ for $\Amod>2$. 

The behavior of $\delta$ with $\eta$ can be qualitatively understood
as follows. For a very massive perturber, the kinetic energy of the
incident flow along the symmetry axis is almost entirely converted to the 
thermal energy of the postshock flow that supports a hydrostatic 
envelope against the gravitational potential of the perturber. 
When the shock is strong ($\mach\gg1$), the postshock 
thermal energy proportional to $\cs^2\mach^2$ balances the gravitational
potential energy $-GM_p/\delta$ at the shock location, resulting
in $\delta/\rs \propto \A/\mach^2$. In the limit of $\mach\rightarrow1$,
the flow does not in principle produce a shock, corresponding to
$\delta \rightarrow \infty$. In practice, the gravitational acceleration
is able to turn an incident nearly-transonic gas into a weakly supersonic 
flow, but the shock that may form is located very far 
away from the perturber anyway. In view of the detached 
shock distance, $\eta$ may be a better indicator of nonlinearity than $\A$;
for fixed $\A$, the shape of a density wake due to a supersonic perturber 
becomes similar to the linear counterpart as $\mach$ increases.

It is well known from hydrodynamic experiments and corresponding theories 
that a supersonic flow over a solid object with a blunt nose 
develops a detached bow shock when the nose angle is larger than
the maximum angle allowed by the postshock flow 
(e.g., \citealt{lie57}, and references therein).
The shocked gas becomes subsonic and slowly adjusts its velocity 
as it flows downstream to meet the boundary conditions at the 
surface of the object. 
In our simulations, the incoming flow toward a massive perturber 
recognizes the hydrostatic envelope as a spherical obstacle. 
Since the nose angle of a 
spherical body with respect to the incident flow is 90 degrees, the
shock must be detached. 

Even though near-hydrostatic envelopes that form in our simulations 
are not entirely impenetrable, 
we want to measure their effective sizes as perturbing obstacles.
This can be achieved by comparing our numerical results for the detached 
shock distances with those from non-gravitating hydrodynamic theories
(or lab experiments).
Assuming that a bow shock ahead of a spherical body with radius $R_s$
has a spherical shape near the symmetry axis,
\citet{guy74} showed that the standoff distance of the shock is 
approximately given by
\begin{equation}\label{eq:guy}
\frac{\delta}{R_s} 
 = \left[\frac{(\gamma-1)\mach^2+2}{4(\mach^2-1)} + 1\right]^{1/(2K)} -1,
\end{equation}
where
\begin{equation}
\frac{1}{K(\mach)} = \frac{1}{2} 
\left[ 1 + \frac{2}{\gamma+1}\frac{1-\mu^2}{\mu}\right]
\left[ 2\mu + 1 + \frac{1}{\mach^2}\right],
\end{equation}
with $\mu^2\equiv ((\gamma-1)\mach^2 +2 )/ (2\gamma\mach^2 -\gamma +1 )$.
\Eq{eq:guy} has proven to explain the experimental data quite well
(\citealp[e.g.,][]{heb50,sch56,van58}; see also \citealt{sch82}).
\Fig{fig:dM_rK} plots as a solid line $\delta/R_s$ from \eq{eq:guy} with 
$\gamma=5/3$ as a function of $\mach$. Also plotted as various symbols
are our numerical results for the ratio of $\delta$ to the BHL radius, 
$\rBHL\equiv GM_p/(V_p^2+\cs^2)$ for models with $\Amod>2$, averaged over 
$t/\tcr>50$. Again, errorbars indicate the standard deviations. 
A rough agreement between the two results suggests that 
the BHL radius can be a useful measure of the effective size
of the hydrostatic sphere.

\subsection{Gravitational Drag Force} \label{sec:force}

For a given wake distribution $\rho(\vct{x},t)$ at time $t$ of 
a perturber, it is straightforward to calculate the DF force 
on it by performing integration in \eq{eq:f_integ}.
\Fig{fig:ftime} plots temporal changes of the DF forces normalized
by $4\pi\rho_\infty (GM_p/\cs)^2$ for models with differing $\A$, 
but with fixed $\mach=1.5$, over the course of the wake evolution.
The dotted line corresponding to the linear DF force 
(\eqo{eq:f_pt_linear}), with $\rmin$ given in \eq{eq:rmin}, 
closely follows the numerical
results for $\A=0.01$, showing that the linear drag force increases 
logarithmically with time. The drag forces for nonlinear cases with 
high $\A$ also have a similar logarithmic time dependence, although 
they fluctuate for a while in response to the 
oscillations of primary vortices as well as detached bow shocks 
before a quasi-steady state is attained;
the fluctuation amplitudes are typically $\sim 4-16\%$, with a smaller
value corresponding to larger $\A$ and $\mach$.
Note that for $\mach=1.5$, the normalized drag force decreases with 
increasing $\A$, indicating that the nonlinear effect makes the DF 
force smaller than the linear estimate. 
This is because a higher value of $\A$ implies a 
correspondingly larger detached shock distance, and 
a hydrostatic envelope with front-back symmetry near the perturber
contributes a negligible amount to the net DF force.

The dimensionless drag forces at $t/\tcr=600$ when the wakes are 
in a quasi-steady state are given in \Fig{fig:fM} for various 
models with different $\A$ and $\mach$. 
For all the supersonic models, the DF force is a decreasing function 
of $\A$. The reduction of the normalized DF force is larger for 
models with $\mach\sim1$ than highly supersonic models. 
The Mach number corresponding to the maximum drag force shifts 
from unity to $\sim1.5$ as the wake becomes highly nonlinear. 
For subsonic models, on the other hand, the nonlinear drag forces
for $\A=20$ and 50
show some fluctuations (represented by errorbars) associated
with slowly-evolving vortices present in the wakes, but their respective
time-averaged values are very close to the drag forces in the linear regime.
In fact, the similarity between the linear and nonlinear drag forces
on subsonic perturbers is expected since the linear density wakes also 
possess front-back symmetry in the vicinity of the perturber
\citep{ost99}.
Since the subsonic DF forces are dominated by the far field where 
perturbations are weak regardless of the perturber mass, the linear
results should be valid even for very massive perturbers.

The gravitational drag force certainly depends on both $\A$ and $\mach$, but 
the discussions given above suggest that it may be through the
nonlinearity parameter $\Amod$.
To check this, we plot in \Fig{fig:fA} the ratio of the nonlinear DF force
$F$ to the linear prediction $F_{\rm lin}$ as a function of $\Amod$. 
Again, various symbols give temporal averages of
$F/F_{\rm lin}$ over $t/\tcr>50$, and 
their standard deviations are indicated by errorbars. For $\Amod\la0.7$ 
with which a bow shock that barely forms is attached to a perturber, 
$F/F_{\rm lin}\approx1$. When $\Amod$ is increased to $\sim 0.7$--2,
the shock becomes detached, but its standoff distance is not so large.
In this case, most of the material in the wake is still located behind of,
but closer to the perturber in comparison with the linear wake
(see, e.g., \Figoabc{fig:compAdens}{b}), 
resulting in the DF force slightly larger
(by less than 20\%) than the linear counterpart. 
In highly nonlinear cases with $\Amod>2$,
however, the presence of a large hydrostatic envelope makes the drag force 
reduced considerably. For $\Amod>2$, the numerical results are well
fitted by 
\begin{equation}\label{eq:fnon}
  F=F_{\rm lin}\left(\frac{\Amod}{2}\right)^{-0.45}.
\end{equation}
\Fig{fig:fA} also plots the gravitational drag force from the hydrodynamic 
simulations of \citet{shi85} for BHL accretion flows, which are consistent 
with our numerical results. We defer to \S\ref{sec:diss} a more detailed 
discussion of our results in connection with the BHL flows.

To ascertain that the reduction of the DF force in highly nonlinear
supersonic cases is really caused by the presence of spherically-symmetric
hydrostatic envelopes near the perturbers, we calculate the drag force
by imposing a cut-off radius $\rmin$ such that only the
regions in the wake with $r>\rmin$ participate in the force 
evaluation (\eqo{eq:f_integ}). 
\Fig{fig:frmin} plots the resulting
dependence of $F$ upon $\rmin$ for the model with $\A=20$ and $\mach=1.5$
at $t/\tcr=600$. 
The vertical dashed line marks the location of the bow shock along the
symmetry axis, while the dotted line indicates a slope of $-1$.
Note that the drag force is independent of $\rmin$ for $\rmin < \delta$,
clearly demonstrating that the hydrostatic sphere
surrounding the perturber has a negligible contribution to the net drag force.
\Fig{fig:frmin} shows $F\propto -\ln(\rmin/\rs)$ for large $\rmin$,
analogous to the linear cases (see \eqo{eq:f_pt_linear}).
From the study of BHL accretion flows to large gravitating bodies,
\citet{sha93} similarly found that the drag force declines logarithmically
as the size of the accretor increases.

\subsection{Resolution Dependency}\label{sec:res}

Finally, we remark on the effect of numerical resolution on the DF force.
\Fig{fig:res} plots the time evolution of the detached shock distance as 
well as the drag force on a perturber with $\A=10$ and $\mach=1.5$ from
models with different resolution: $\Delta z/\rs = 0.1$, 0.2, 0.4, and 0.8,
where $\Delta z$ is the grid spacing.
In models with $\Delta z/\rs = 0.4$ and 0.8, the shock has a larger 
standoff distance without noticeable oscillations and the DF
force is correspondingly smaller 
than in models with higher resolution.
Compared with the $\Delta z/\rs=0.2$ model, the model with $\Delta z/\rs=0.1$ 
presents the shock oscillations with higher amplitudes and 
arrives at quasi-steady equilibrium earlier. 
This resolution dependence of 
the shock oscillations was also noticed by \citet{mat89} for adiabatic
BHL flows onto an absorbing perturber.
The resolution study shown in \Fig{fig:res} suggests that our numerical 
results for the DF forces are reliable as long as 5 or more grids 
per $\rs$ are taken. 

\section{SUMMARY AND DISCUSSION}\label{sec:diss}

DF of bodies orbiting in a gaseous medium is of great importance in various
astronomical systems ranging from protoplanetary disks to galaxy clusters.
In previous analytic studies of DF, it has been assumed that the mass of 
a moving object is small enough for the induced density wake to have low 
amplitudes and be thus in the linear regime. However, there are many 
astronomical situations such as in a merger of black holes near a galaxy 
center and migration of protoplanets, where a perturber is so massive 
that the induced wakes are well in the nonlinear regime.
In this paper, we use numerical hydrodynamic simulations to explore
nonlinear gravitational responses of the gas to, and the resulting
drag force on, a perturber with mass $M_p$ moving straight at velocity $V_p$ 
in an initially static, uniform background with density $\rho_\infty$ and 
adiabatic sound speed $\cs$. The perturber is represented by a Plummer sphere
with softening radius $\rs$. Unlike in the BHL problems, the perturber
in our models does not possess a defined surface through which gas is
accreted or reflected.
Assuming an axial symmetry, we solve the basic equations on
the ($R$, $z$) plane, where $R$ and $z$ denote the direction perpendicular
and parallel to the motion of the perturber, respectively.
Our numerical models are completely
characterized by two dimensionless parameters:
$\A=GM_p/(\cs^2\rs)$ and $\mach=V_p/\cs$.
To study DF in various situations, we run as many as 58 models that
differ in $\A$ and $\mach$. Our standard models have 5 zones per $\rs$,
but we also run models with different resolutions to ensure that
the density and velocity wakes are fully resolved.

For supersonic models ($\mach>1$), we find that a massive perturber 
with $\A\gg1$
produces a bow shock ahead of it through which the incident supersonic
flow becomes subsonic. 
In the beginning, the postshock flow in the nonlinear cases develops
transient features such as a primary vortex ring and an 
associated counterstream near the symmetry axis, which causes the shock to
oscillate around its equilibrium position.
The shock oscillation added by the counterstream
is overstable, amplifying the amplitudes of its oscillation and 
the vortex movement in the ($R$, $z$)-plane (see \S\ref{sec:general}).
The vortex ring is eventually shed downstream from the perturber,
leaving the nearly-stationary bow shock and a quasi-hydrostatic 
envelope that surrounds the perturber.
On the other hand, subsonic models with $\A\gg1$ without involving a shock
retain a primary vortex and many other small-scale structures 
until the end of runs. Nevertheless, strong gravity makes 
the density distribution near the perturber still very close to that 
under hydrostatic equilibrium. 

By comparing the numerical results from various supersonic models with 
differing $\A$ and $\mach$, we find that the simulation outcomes such 
as the detached shock distance and the gravitational DF force are very
well quantified by a single parameter $\eta$ defined in \eq{eq:Amod}.
When $\Amod\la0.7$, the system is in the linear regime where
a Mach cone is bounded by low-amplitude Mach waves, 
and the drag force is just the same as the analytic linear value 
$\Fplin$ of \citet{ost99}. When $\Amod$ is moderate at $0.7\la\Amod\la2$,
the Mach waves turn into a bow shock that is weakly detached,
with the standoff distance $\delta \approx 2(\Amod/2)^{2.8}\rs$
from the perturber. In this case, the density wake is slightly shifted 
toward the perturber compared with the linear counterpart, causing the 
drag force to be larger than $\Fplin$ by a factor of less than 1.2. 
In the highly nonlinear regime with $\Amod>2$, however, the detached 
shock distance behaves as $\delta=\eta\rs$, and the nonlinear drag force 
$F$ is given by $F/\Fplin=(\eta/2)^{-0.45}$.
The reduction of the drag force compared with the linear value is
because the density wake close to the perturber is in near-hydrostatic
equilibrium and thus contributes a negligible amount to the net
DF force. Since front-back symmetry notable for nonlinear wakes
exists also in the linear wakes with $\mach<1$, the nonlinear drag force
on a massive subsonic perturber is similar to the linear prediction.

As mentioned in \S\ref{sec:intro}, our model simulations differ
from those of the BHL accretion flows in terms of the boundary conditions.
In our models, a perturber simply provides a gravitational potential for
the background gas and does not hold any surface, while models for the
BHL accretion considered a defined surface through which the gas is accreted or
reflected. The different boundary conditions might lead to different
evolution and structure of wakes. For instance, the BHL accretion flows 
around a perturber present nonlinear features including a detached bow shock, 
vortex oscillations, and counterstreams
in the forward/backward directions, just as in our models,
if the perturber is not perfectly absorbing
\citep{shi85,fry87,mat89}. 
When the perturber has a totally absorbing surface, on the other hand,
the wakes are relatively quiescent, flowing nearly spherically into the 
perturber that absorbs the angular momentum carried by the accreting gas
\citep{shi85,fry87,koi91,ruf94}.
In the case of an extremely small perturber compared with the accretion radius,
even an absorbing body produces violent features since it is not able to 
accept the whole accreting gas \citep{koi91}.

Despite these differences, the detached distances of bow shocks and
the resulting DF forces appear to be not so sensitive to the boundary
conditions adopted.
\Fig{fig:dEtaBHL} plots the detached shock distances inferred from 
the results of two-dimensional axisymmetric simulations 
\citep{shi85, fry87,mat89,koi91} 
and the tabulated values from three-dimensional nonaxisymmetric runs 
\citep{ruf94} for the BHL accretion flows,
by taking $\rs$ equal to the radius of the perturber.
The BHL results under the reflecting boundary conditions are larger than
our results by a factor of only $\sim1.2$, while those under the
absorbing boundary conditions are smaller by a factor of $\sim0.4$.
Since the work on the BHL accretion mostly focused on mass accretion rates, 
only a few of them evaluated the gravitational drag forces. 
In \Fig{fig:fA}, we plot using star symbols 
the drag forces from the adiabatic runs with $\gamma=5/3$ in \citet{shi85}, 
which are the only data we acquired from the literature that allow
a reliable comparison with our results\footnote{Although \citet{sha93} and 
\citet{ruf94} also gave the drag coefficients, the perturbers are too
large in the former and the units are uncertain in the latter to be
compared with our results.}.
Note that the drag force is larger on a purely-absorbing perturber for 
which the detached shock distance is smaller. 
The DF forces from \citet{shi85} are roughly consistent with, and 
follow a similar trend to, our results. This is probably because even 
though density wakes are nowhere close to being hydrostatic in BHL accretion
flows, they somehow maintain front-back symmetry that makes the net DF force 
smaller.

While we have considered in this paper only an adiabatic gas with index 
$\gamma=5/3$, we note that the DF force may depends sensitively on $\gamma$. 
Simulations of the BHL accretion flows reported that as $\gamma$ decreases,
the detached shock distance decreases and the density wake becomes shaped 
increasingly into a cone-like structure similar to the linear wake 
\citep{hun79,shi85,ruf95,ruf96}.
This is presumably because bow shocks in lower-$\gamma$ models 
should be stronger in order to compensate for the diminished pressure 
in supporting the postshock gas against the perturber's gravity.
Consequently, drag forces are larger for the gas with smaller $\gamma$
\citep{shi85}.

\citet{esc05} carried out numerical simulations of a DF-induced merger of 
supermassive binary black holes due to an adiabatic gas with $\gamma=5/3$.
Their Figure~9 presents the effect of the black hole mass 
on the evolution of the binary separation. From this figure, we infer 
that the orbital decay time scales with the black hole mass as 
$\tau_{\rm decay}\propto M_p^{-0.3}$ in their simulations, which
is somewhat shallower than the linear expectation
of $\tau_{\rm decay} \propto M_p^{-1}$ (e.g., \citealt{ost99}). 
On the other hand, our results predict $\tau_{\rm decay} 
\propto M_p^{-0.55}$ if the perturber is sufficiently massive, 
suggesting that the delayed orbital decay of supermassive black holes in
\citet{esc05} is partly due to the nonlinear effect.
Of course, there are many other factors that may change $\tau_{\rm decay}$.
While we consider quite an ideal situation in which
a perturber is moving straight in a static, uniform medium, 
the gas in \citet{esc05} is distributed in a rotating, radially-stratified, 
self-gravitating disk and the black holes follow curvilinear, possibly 
eccentric, orbits. 
In addition, the Mach number of the black holes in their models is likely 
to vary during the decay toward the orbital center, which may also 
modify the decay time. Since these compact objects have 
a very small size, it is also an issue whether numerical models 
resolve detached bow shocks, which is crucial in evaluating the drag
force accurately. We will discuss these in subsequent work.

\acknowledgments
We acknowledge a helpful report from an anonymous referee and
Eve C.\ Ostriker for useful suggestions.
The software used in this work was in part developed by the DOE-supported
ASC/Alliance Center for Astrophysical Thermonuclear Flashes at the University
of Chicago.
This work was supported by KICOS through the grant
K20702020016-07E0200-01610 provided by MOST.
Simulations were performed by using the supercomputing resource of the 
Korea Institute of Science and Technology Information through 
the grant KSC-2009-S02-0008.

\appendix
\setlength\arraycolsep{2pt}
\section{Linear Solution for the Velocity Wake}
\label{sec:FT}

Using the Fourier transform for the space variables and the Laplace
transform for the time variable, 
\citet{just90} derived an analytic solution for the velocity
field in the linear wake, but unfortunately the resulting
expression (their equation~[47]) contains an typographical mistake. 
Here we rederive the expression 
using Fourier transforms for both space and time variables.
We consider a point-mass perturber with mass $M_p$ on 
a straight-line trajectory through a uniform medium with density
$\rho_\infty$ and adiabatic sound speed $\cs$. 
The basic equations of hydrodynamics for an inviscid, 
nonmagnetic, non-self-gravitating gas are
\begin{equation}\label{eq:con}
  \frac{\partial\rho}{\partial t}+\vct{\nabla}\cdot(\rho\vct{v})=0,
\end{equation}
and
\begin{equation}\label{eq:mom}
  \frac{\partial\vct{v}}{\partial t}+\vct{v}\cdot\vct{\nabla}\vct{v}
  = -\frac{1}{\rho}\vct{\nabla}P-\vct{\nabla}\Phi_\ext,
\end{equation}
where $\Phi_\ext$ is the gravitational potential of the perturber 
with density $\rho_\ext$, satisfying
the Poisson equation $\vct{\nabla}^2\Phi_\ext=4\pi G\rho_\ext$. 
Assuming that the density and velocity fields induced by the perturber 
have small amplitudes, we linearize \eqs{eq:con}{eq:mom} to obtain
\begin{equation}\label{eq:con_p}
  \frac{\partial\alpha}{\partial t}+\vct{\nabla}\cdot\vct{v}=0,
\end{equation}
and
\begin{equation}\label{eq:mom_p}
  \frac{\partial\vct{v}}{\partial t} + 
  \cs^2\vct{\nabla}\alpha = -\vct{\nabla}\Phi_\ext,
\end{equation}
where $\alpha \equiv \rho/\rho_\infty-1$ is the dimensionless
perturbed density.

Taking Fourier transformation of \eqs{eq:con_p}{eq:mom_p}, we have 
\begin{equation}\label{eq:FTd}
  \left(k^2-\frac{\omega^2}{\cs^2}\right)\widehat{\alpha}
  =\frac{4\pi G}{\cs^2}\widehat{\rho}_\ext
\end{equation}
and
\begin{equation}\label{eq:FTv}
  \widehat{\vct{v}}=\frac{\omega\vct{k}}{k^2} \widehat{\alpha},
\end{equation}
where symbols with and without hat denote a Fourier pair defined by
\begin{equation}\label{eq:IFT}
  \Psi(\vct{x},t)=\frac{1}{(2\pi)^4}\int\!\!\!\!\int\!\!\!\!\int\!\!\!\!\int
  \widehat{\Psi}(\vct{k},\omega)e^{i(\vct{k}\cdot\vct{x}-\omega t)}
  \,d^3\vct{k}\,d\omega
\end{equation}
and
\begin{equation}\label{eq:FT}
  \widehat{\Psi}(\vct{k},\omega)=\int\!\!\!\!\int\!\!\!\!\int\!\!\!\!\int
  \Psi(\vct{x},t)e^{-i(\vct{k}\cdot\vct{x}-\omega t)}\,d^3\vct{x}\,dt,
\end{equation}
with $\Psi$ representing any physical quantities.

For a perturber introduced at $t=0$ and moving at a constant
speed $V_p$ along the positive $z$-direction,
$\rho_\ext=M_p\,\delta(\vct{x}-V_pt\ez)\,\step(t)$, where $\delta(\vct{x})$
is the Kronecker delta and $\step(t)$ is a Heaviside step function.
This results in 
\begin{equation}\label{eq:FTd_ext}
  \widehat{\rho}_\ext= 
  M_p\left[2\pi\delta(\omega-k_z V_p)+\frac{i}{\omega-k_z V_p}\right].
\end{equation}
The second term in \eq{eq:FTd_ext} is originated from the
step function, which vanishes if a steady state is assumed
\citep[e.g.,][]{fur02}.

Substituting \eq{eq:FTd_ext} into \eq{eq:FTd},
and taking an inverse Fourier transform of the resulting
expression for $\widehat{\alpha}$, one can obtain the solution
of the perturbed density for finite-time perturbations. 
The detailed procedure is similar to
that of the Fourier-Laplace transform method of \citet{just90},
so that we simply write the result
\begin{equation}\label{eq:alpha}
  \alpha=\frac{GM_p}{\cs^2}\frac{\Upsilon_1}{(s^2+R^2(1-\mach^2))^{1/2}},
\end{equation}
where $s\equiv z-V_pt$ and $\Upsilon_1 =2$, 1, and 0 in
region I ($R^2+z^2>\cs^2t^2,\ R^2<-sz,\ s^2+R^2(1-\mach^2)>0$, $\mach>1$),
region II ($R^2+z^2<\cs^2t^2$), and any other region, respectively.
Note that \eq{eq:alpha} is identical to equation~(10) of \citet{ost99} 
based on the retarded Green's function technique.

Combining \eqthree{eq:FTd}{eq:FTv}{eq:FTd_ext} together in favor of 
$\widehat{\vct{v}}$ and taking its inverse Fourier transform, we have
\begin{equation}\label{eq:IFTv}
  \vct{v}=\frac{GM_p}{4\pi^3\cs^2}\int\!\!\!\!\int\!\!\!\!\int\!\!\!\!\int
  \bigg[\frac{2\pi\delta(\omega-k_z V_p)}{k^2-\omega^2/\cs^2}
  +\frac{i}{(k^2-\omega^2/\cs^2)(\omega-k_z V_p)}\bigg]
  \frac{\omega\vct{k}}{k^2} 
  e^{i(\vct{k}\cdot\vct{x}-\omega t)}\,d^3\vct{k}\,d\omega.
\end{equation}
The second term in the integrand of \eq{eq:IFTv} has
three simple poles on the real $\omega$ axis at $\omega_1=k_z V_p$
and $\omega_{2,3}=\pm k\cs$. To ensure the proper analyticity, we must
displace them by an infinitesimal amount into the upper or lower half 
$\omega$-plane. Since our problem requires $\vct{v}=0$ everywhere 
for $t<0$, the proper choice of the pole displacements should be
$\omega_1$ into the upper half $\omega$-plane
and $\omega_{2,3}$ to the lower half $\omega$-plane.
For $t<0$, then, a contour of integration consisting of the real
$\omega$ axis and an infinitely-large semicircle in the upper
half $\omega$-plane encloses only $\omega_1$, and the calculus 
of residues guarantees that after the integration over $\omega$, 
the first and second terms in the integrand of \eq{eq:IFTv} 
become identical with each other with opposite sign, resulting in 
no velocity perturbations before the introduction of the perturber.

For $t>0$, the contour that goes along an infinitely-large semicircle 
in the lower half $\omega$-plane instead contains the poles at 
$\omega_{2,3}$. The remaining steps in the integration with respect to 
$\omega$ is quite straightforward to yield 
\begin{equation}\label{eq:IFTv_a}
  \vct{v}=\frac{GM_p}{4\pi^2\cs}\int\!\!\!\!\int\!\!\!\!\int
  \bigg(
  \frac{e^{ik\cs t}-e^{-ik_zV_pt}}{k+k_z\mach} -
  \frac{e^{-ik\cs t}-e^{-ik_zV_pt}}{k-k_z\mach} 
  \bigg) 
  \frac{\vct{k}}{k^2} e^{i\vct{k}\cdot\vct{x}}\,d^3\vct{k}.
\end{equation}
To carry out the integration in \eq{eq:IFTv_a}, 
it is convenient to use spherical coordinates in the $\vct{k}$-space as
\begin{equation}\label{eq:k_spher}
  \vct{k}=k\,
  (\sin\theta\cos(\varphi+\zeta), 
  \sin\theta\sin(\varphi+\zeta), 
  \cos\theta),
\end{equation}
which is oriented relative to the space coordinates
\begin{equation}\label{eq:x_cylin}
  \vct{x}=(R\cos\zeta,R\sin\zeta,z), 
\end{equation}
satisfying $\vct{k}\cdot\vct{x}=k\,(z\cos\theta+R\sin\theta\cos\varphi)$.
We then obtain
\begin{eqnarray}\label{eq:v_int3} 
  \vct{v}=\frac{GM_p}{4\pi^2\cs}\int\!\!\!\!\int\!\!\!\!\int
  \bigg[
   \frac{e^{ik(z\cos\theta+\cs t)} - e^{iks\cos\theta}}{1+\mach\cos\theta}
  -\frac{e^{ik(z\cos\theta-\cs t)} - e^{iks\cos\theta}}{1-\mach\cos\theta}
  \bigg] e^{ikR\sin\theta\cos\varphi}\nonumber\\
  \times
  (\sin^2\theta\cos(\varphi+\zeta)\ex
  +\sin^2\theta\sin(\varphi+\zeta)\ey
  +\sin\theta\cos\theta\ez) \,\,dk\,d\theta\,d\varphi,
\end{eqnarray}
where $\ex$, $\ey$, and $\ez$ are the unit vectors in the $x$-, $y$-, and 
$z$-directions, respectively. Without loss of generality, we take $\zeta=0$, 
in which case $\ex$ and $\ey$ correspond to the unit vectors in the radial
($\eR$) and azimuthal ($\vct{\mathrm{e_\varphi}}$) directions, respectively. 
It then follows that the azimuthal velocity vanishes (i.e., $v_\varphi=0$)
since this component in the integrand is an odd function of $\varphi$.

For the velocity component in the direction of motion, 
we substitute $\theta=\pi-\theta^\prime$ to the second term 
inside the square bracket of \eq{eq:v_int3} to obtain
\begin{equation}\label{eq:vz_int2}
  v_z=\frac{GM_p}{\pi\cs}
    \int^\pi_0\!\!\!d\theta\!\int^\infty_0\!\!\!dk\,
    \frac{\sin\theta\cos\theta J_0(kR\sin\theta)}{1+\mach\cos\theta}
  \left\{
  \cos[k(z\cos\theta+\cs t)]
  -\cos(ks\cos\theta)
  \right\}.
\end{equation}
Using [G6.671.8]\footnote{[G$\cdots$] refers to the number of formula 
in the integral table of \citet{itable}.}, inserting the substitutions
\begin{equation}\label{eq:subst2}
  \sin\eta=\frac{(R^2+z^2)\cos\theta+z\cs t}{R(R^2+z^2-\cs^2 t^2)^{1/2}}
\end{equation}
and
\begin{equation}\label{eq:subst1}
  \sin\eta=\frac{(R^2+s^2)^{1/2}}{R}\cos\theta,
\end{equation}
for the first and second terms of \eq{eq:vz_int2}, respectively, 
and applying [G2.551.3] for the $\eta$-integrations, we obtain
\begin{equation}\label{eq:vz}
  v_z=-\frac{GM_p}{V_p}
  \left[
   \frac{1}{(R^2+s^2)^{1/2}}
  -\frac{\Upsilon_1}{(s^2+R^2(1-\mach^2))^{1/2}}
  -\frac{\Upsilon_2}{(R^2+z^2)^{1/2}}
  \right],
\end{equation}
where $\Upsilon_1=$2, 1, 0 and $\Upsilon_2=1$, 0, 1
for region I ($R^2+z^2>\cs^2t^2,\ R^2<-sz,\ s^2+R^2(1-\mach^2)>0$,
$\mach>1$),
region II ($R^2+z^2<\cs^2t^2$),
and any other region, respectively.

For the velocity component in the radial direction,
the use of [G6.671.1] similarly leads to
\begin{equation}\label{eq:vR}
  v_R=\frac{GM_p}{RV_p}
  \left[
   \frac{s}{(R^2+s^2)^{1/2}}
  -\frac{s\Upsilon_1}{(s^2+R^2(1-\mach^2))^{1/2}}
  -\frac{z\Upsilon_2}{(R^2+z^2)^{1/2}}
  \right].
\end{equation}
The results of our time-dependent linear analyses for density and 
velocity fields are confirmed 
by direct numerical simulations for a low-mass perturber 
with $\A=0.01$ presented in \S\ref{sec:linear}.

In terms of our notation\footnote{The conversion of symbols used
in \citet{just90} to those in the current paper is
$V_0 \rightarrow -V_p$, $z\rightarrow s$, $z_g \rightarrow z$,
$r_g \rightarrow (R^2 + z^2)^{1/2}$, $r = (R^2 + s^2)^{1/2}$, 
and $c_a \rightarrow \cs$.}, equation (47) of \citet{just90} can
be rewritten as 
\begin{equation}\label{eq:just}
  (v_R, v_z)=\frac{GM_p}{RV_p}
  \left[
   \frac{(s, -R)}{(R^2+s^2)^{1/2}}
  -\frac{(s, -R)\Upsilon_1}{(s^2+R^2(1-\mach^2))^{1/2}}
  -\frac{(z, -R)}{(R^2+z^2)^{1/2}}
  \right].
\end{equation}
Comparison of \eq{eq:just} with \eqs{eq:vz}{eq:vR} shows that 
equation (47) of \citet{just90} takes $\Upsilon_2=1$ everywhere, which
is clearly incorrect.

\clearpage

\begin{figure}
  \epsscale{0.7}
  \plotone{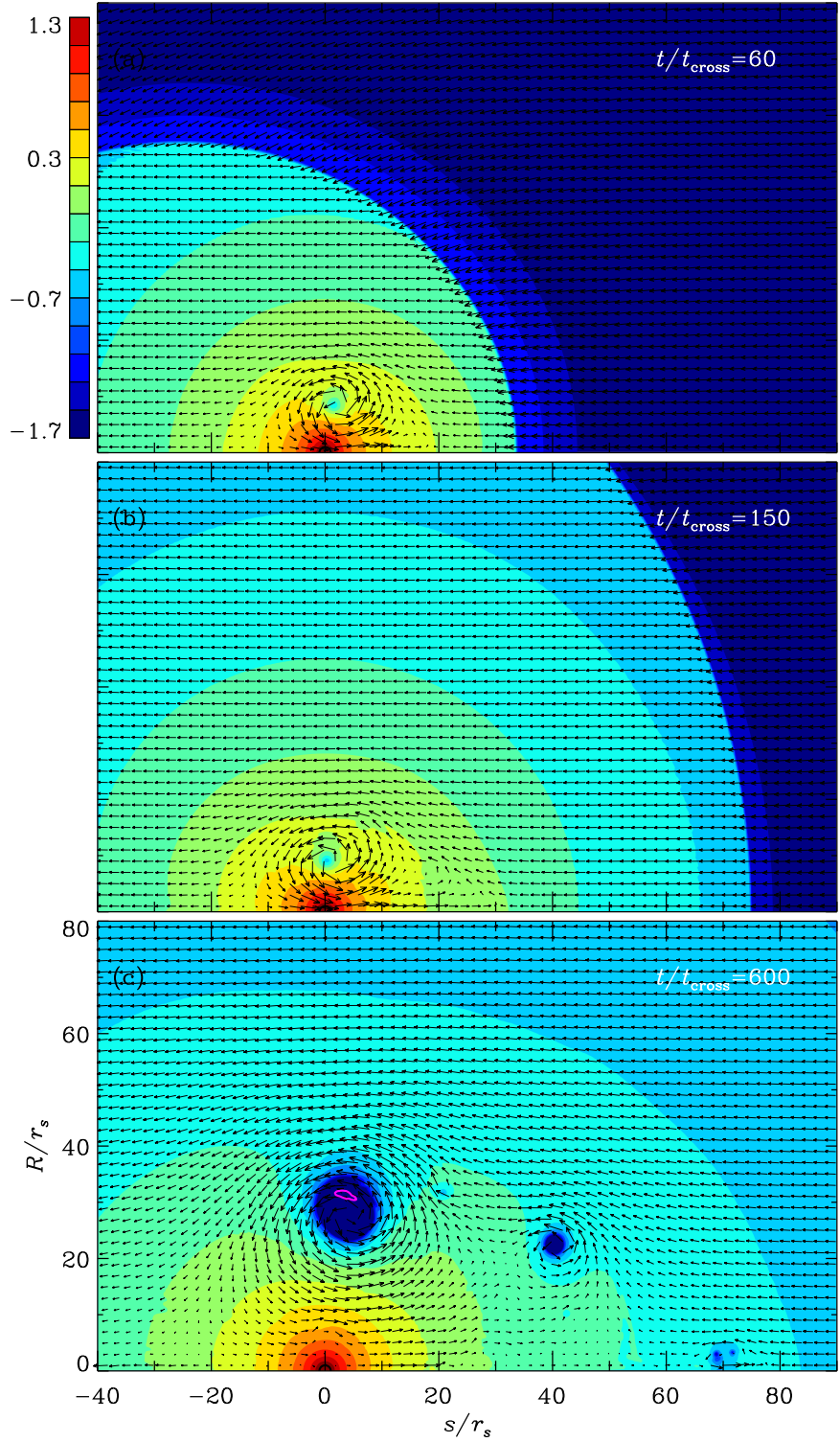}
  \caption{\label{fig:subs}
  Snapshots of the perturbed gas density 
  (logarithmic color-scale) overlaid with the velocity structure
  (arrows) of a nonlinear subsonic model 
  with  $\A=20$ and $\mach=0.5$ in a comoving 
  frame with the perturber.  Colorbar labels 
  $\log (\rho/\rho_\infty-1)$.} 
\end{figure}

\begin{figure}
  \epsscale{1.0}
  \plotone{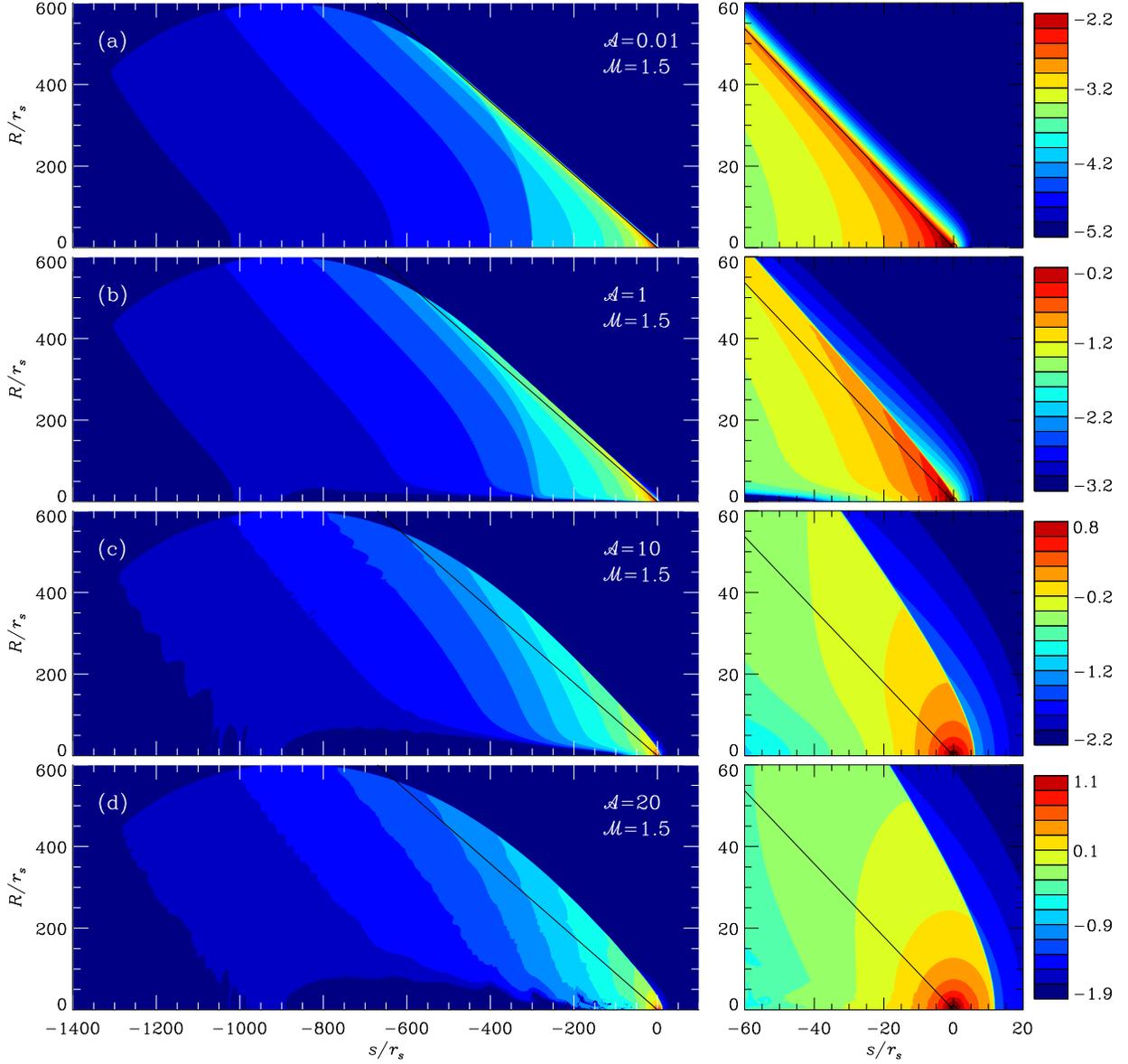}
  \caption{\label{fig:compAdens}
  Distributions of quasi-steady density wakes with varying $\A$ on
  the $s$--$R$ plane at $t/\tcr=600$, 
  with the perturber placed at $(s,R)=(0,0)$. 
  Left panels show large-scale views of $\alpha$, while the right
  panels focus on a small section with $-60 \leq s/\rs \leq 20$
  and $0 \leq R/\rs < 60$ near the perturber.
  All the models have the same $\mach=1.5$. 
  The black line in each panel traces the Mach cone formed by
  a low-mass point-mass perturber with the same Mach number.
  The perturber gathers more gas toward it as $\A$ increases. 
  Colorbars label $\log (\rho/\rho_\infty-1)$. }
\end{figure}

\begin{figure}
  \epsscale{0.8}
  \plotone{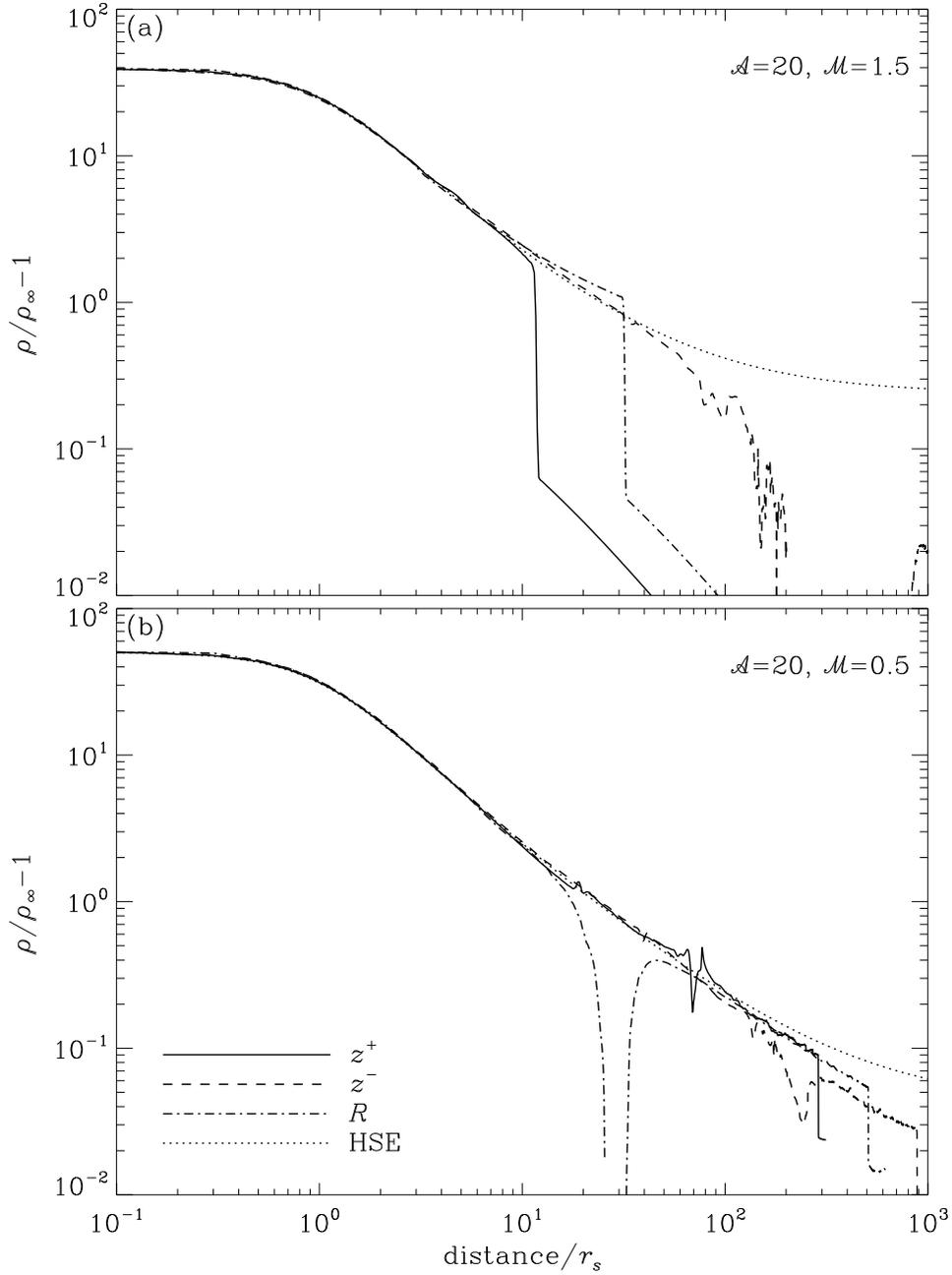}
  \caption{\label{fig:HSE}
  Density profiles along the positive-$z$ (\textit{solid}), 
  negative-$z$ (\textit{dashed}), and $R$ (\textit{dot-dashed}) axes 
  from the perturber center for (\textit{a}) a supersonic model with
  $\A=20$ and $\mach=1.5$ and (\textit{b}) a subsonic model with
  $\A=20$ and $\mach=0.5$ at $t/\tcr=600$. 
  In each panel, the dotted line
  gives the respective density distribution under the assumption of 
  hydrostatic equilibrium, which is in good agreement with the numerical
  results for $r/\rs \la 10$.}
\end{figure}

\begin{figure}
  \epsscale{1.0}
  \plotone{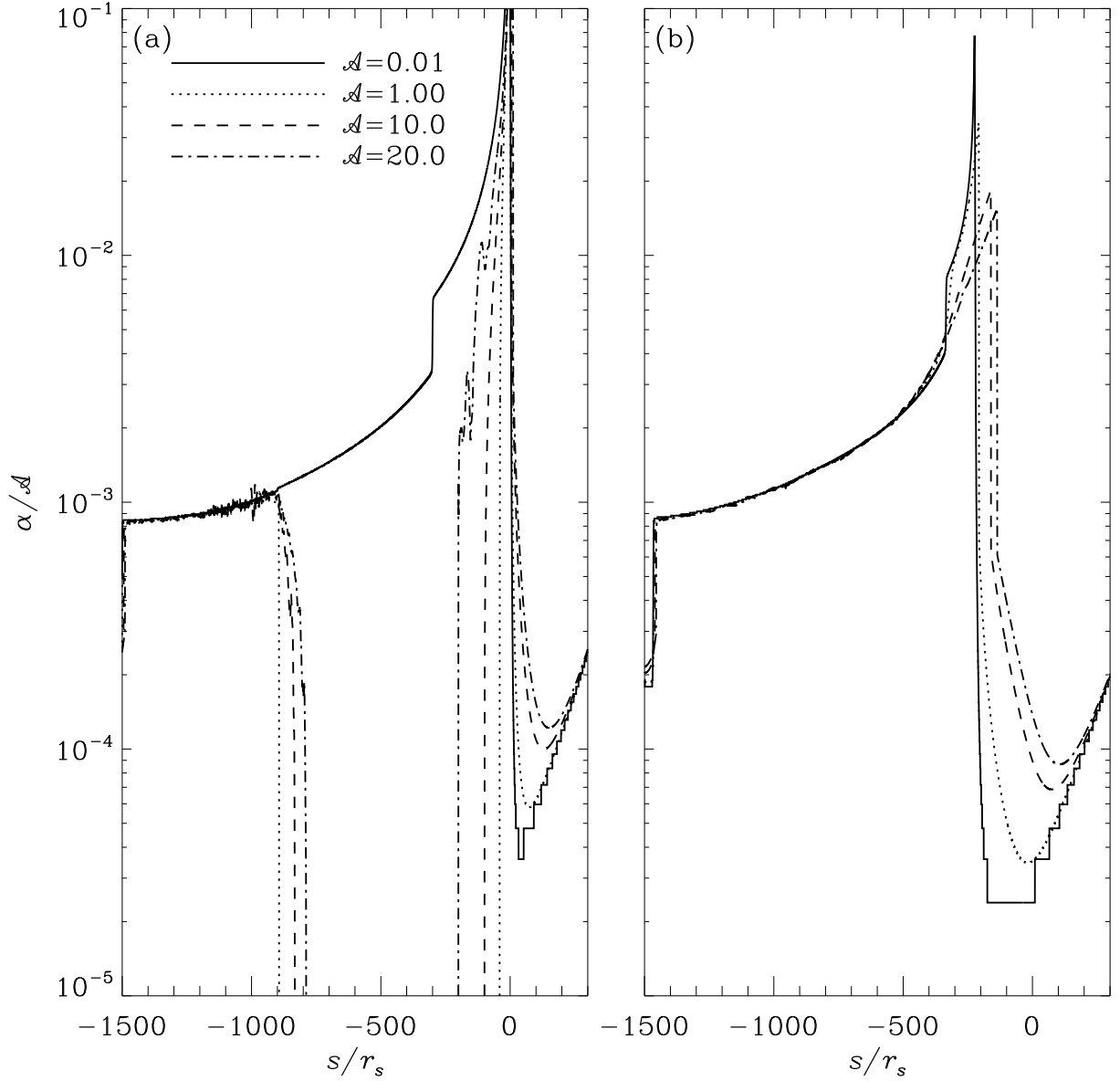}%
  \caption{\label{fig:profiles}
  Distributions of normalized perturbed density $\alpha/\A$ 
  as functions of $s$ along the (\textit{a}) $R=0$ and (\textit{b}) 
  $R/\rs=200$ cuts for models shown in \Fig{fig:compAdens}.
  While $\alpha/\A$ near the perturber with $\A\ge 1$ deviates significantly 
  from that of the linear case, it becomes nearly independent
  of $\A$ in the regions far away from the perturber.
  For $\A\ge 1$, $\alpha$ shows some fluctuations and becomes negative 
  in some regions near the symmetry axis ($R=0$) 
  behind the perturber, since the flow
  in these regions already experienced strong perturbations 
  in the upstream region.  }
\end{figure}

\begin{figure}
  \epsscale{1}
  \plotone{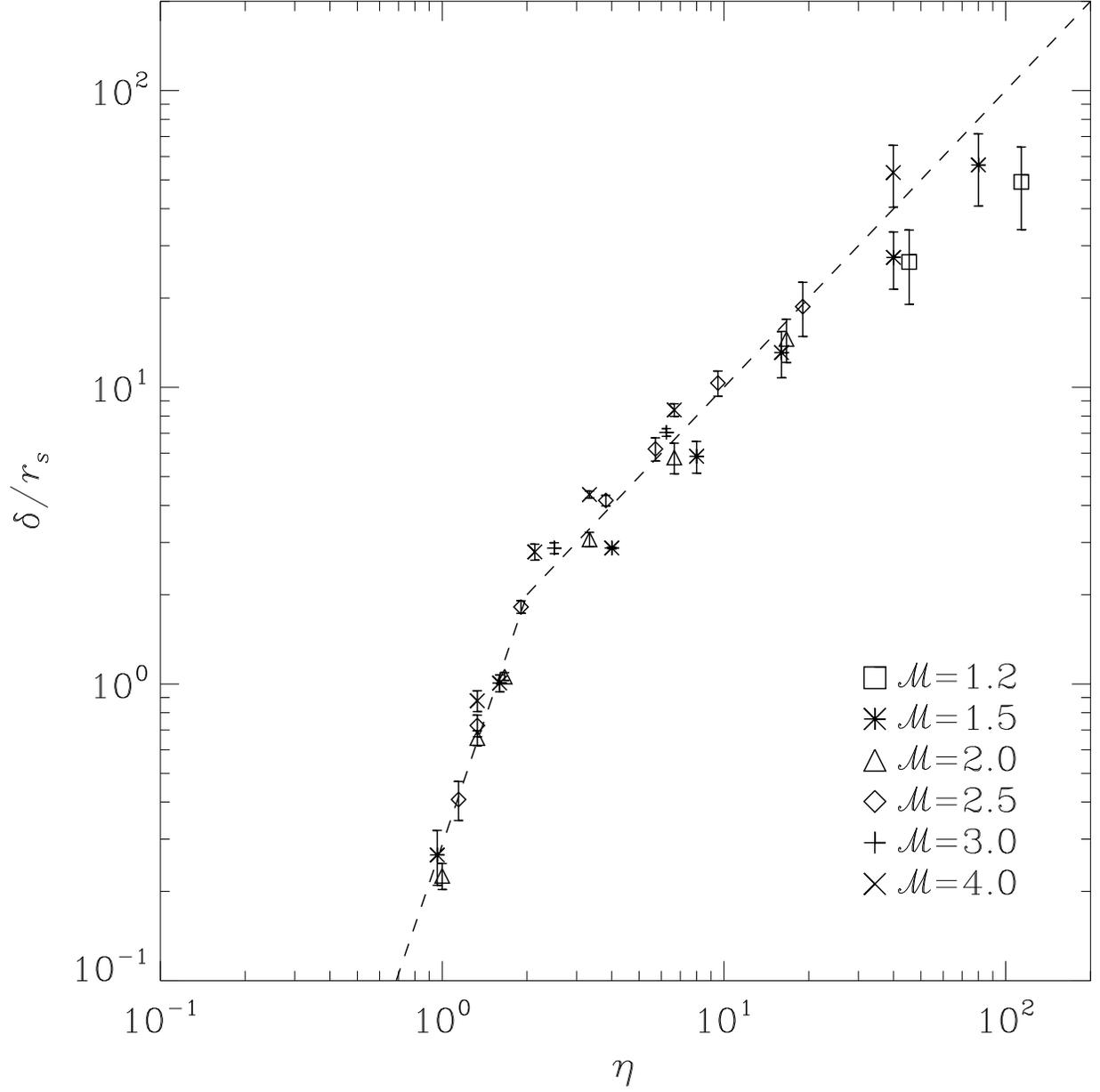}
  \caption{\label{fig:dEta}
  Detached shock distance $\delta$ as a function of the nonlinearity 
  parameter $\Amod=\A/(\mach^2-1)$. 
  Various symbols and errorbars give the means and standard deviations of 
  $\delta$ over time for $t/\tcr>50$. 
  Dashed lines correspond to the broken power laws 
  $\delta/\rs=2(\eta/2)^{2.8}$ for $0.7\la \eta \la 2$ and
  $\delta/\rs=\eta$ for $\eta\ga 2$ that describe the numerical 
  data quite well.}
\end{figure}

\begin{figure}
  \epsscale{1}
  \plotone{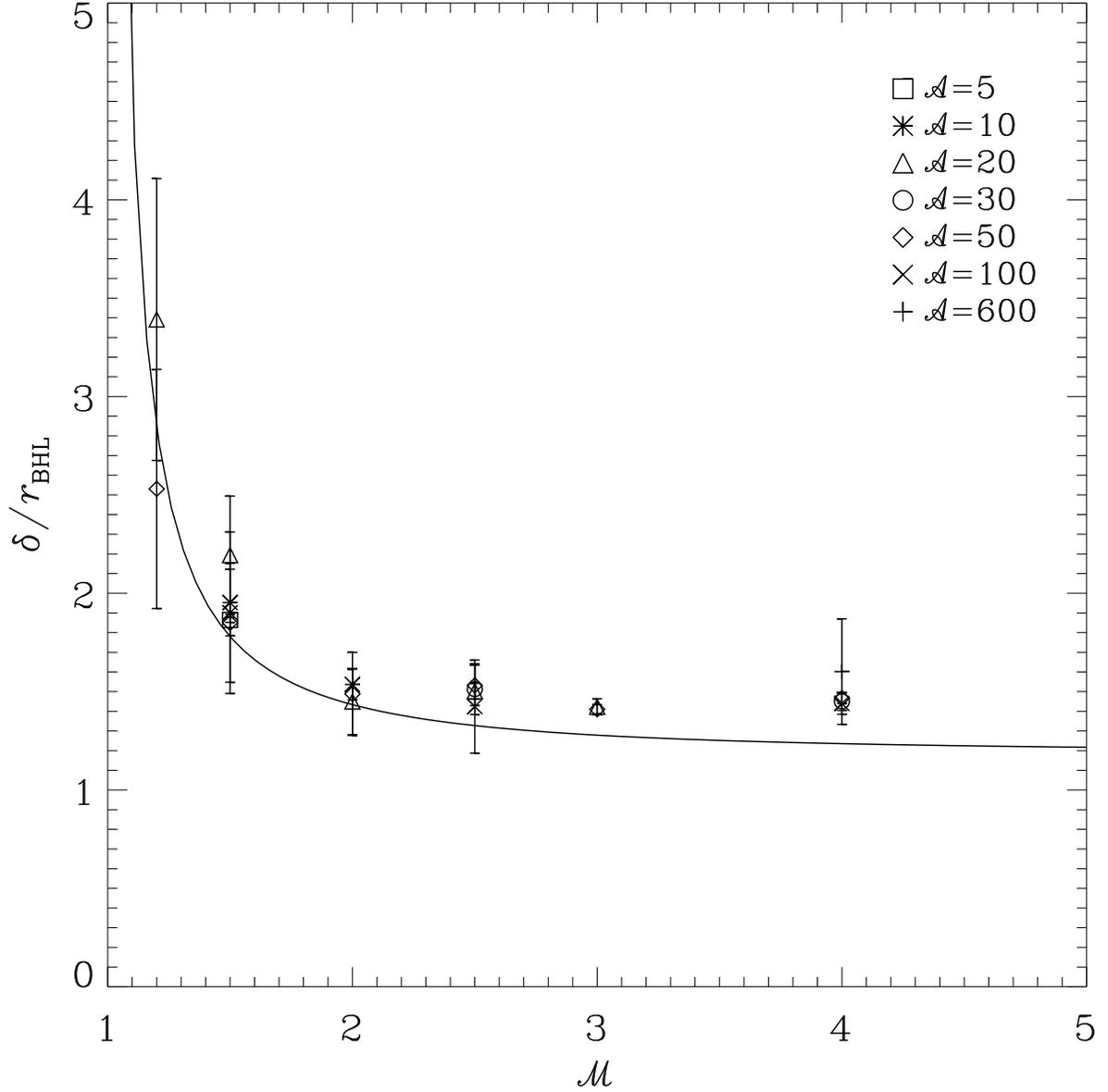}
  \caption{\label{fig:dM_rK}
  Dimensionless detached shock distance as a function of the Mach number.
  Various symbols and errorbars give the means and standard deviations of
  $\delta/\rBHL$ over time for $t/\tcr>50$, where
  $\rBHL$ is the BHL radius.
  The solid curve plots \eq{eq:guy} for $\delta/R_s$, 
  the ratio of the standoff distance of the shock 
  to the radius of a spherical body in the non-gravitating
  hydrodynamic theory.} 
\end{figure}

\begin{figure}
  \epsscale{1}
  \plotone{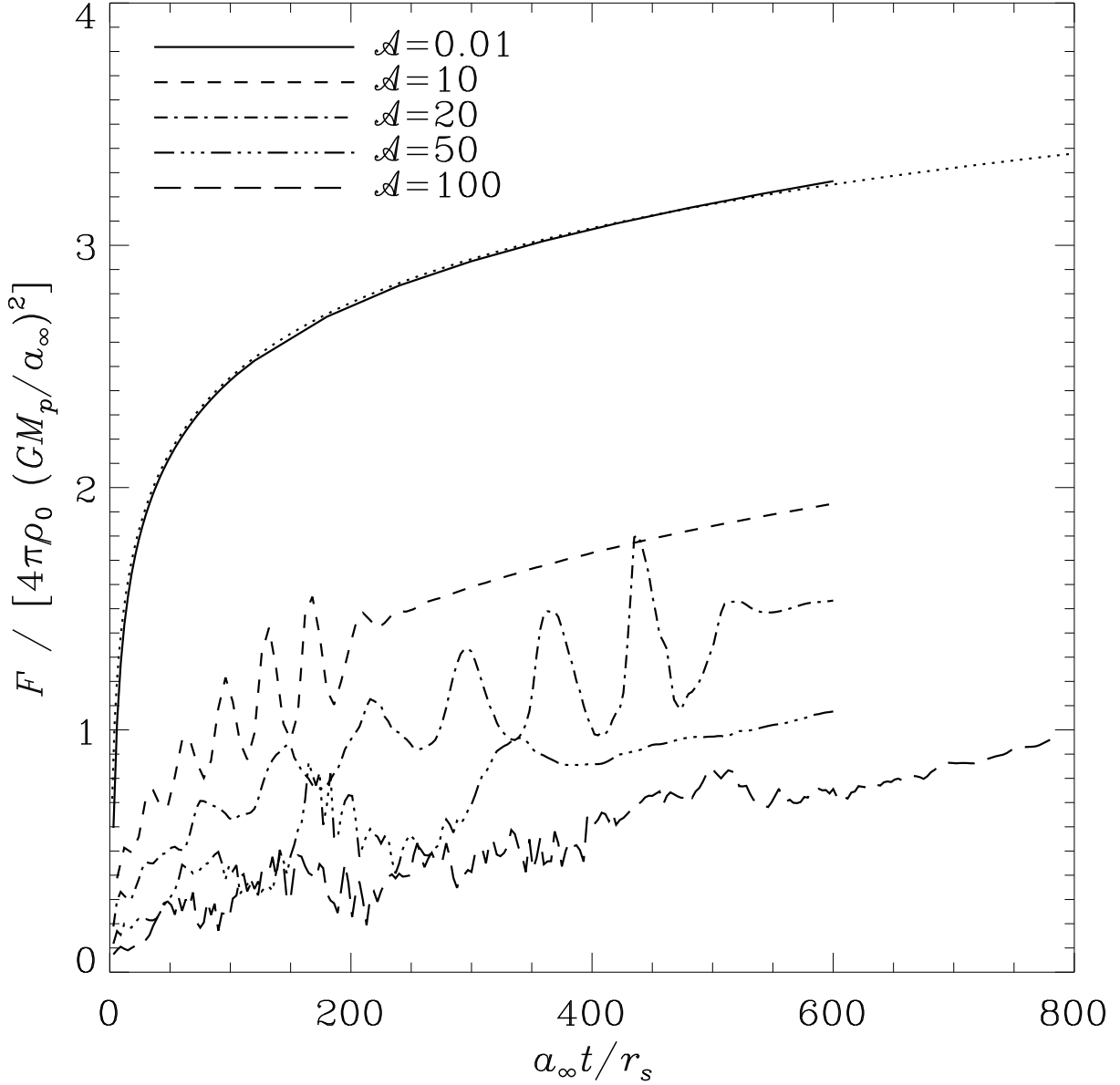}
  \caption{\label{fig:ftime}
  Temporal changes of the dimensionless DF forces for models with $\mach=1.5$. 
  The dotted line is the prediction of \eq{eq:f_pt_linear} with
  $\rmin=0.35\mach^{0.6}\rs$, in good agreement with the
  numerical result for a low-mass perturber with $\A=0.01$. 
  The drag forces increase logarithmically with time, 
  although models with large $\A$ exhibit early fluctuations 
  in accordance with shock oscillations.}
\end{figure}

\begin{figure}
  \epsscale{1}
  \plotone{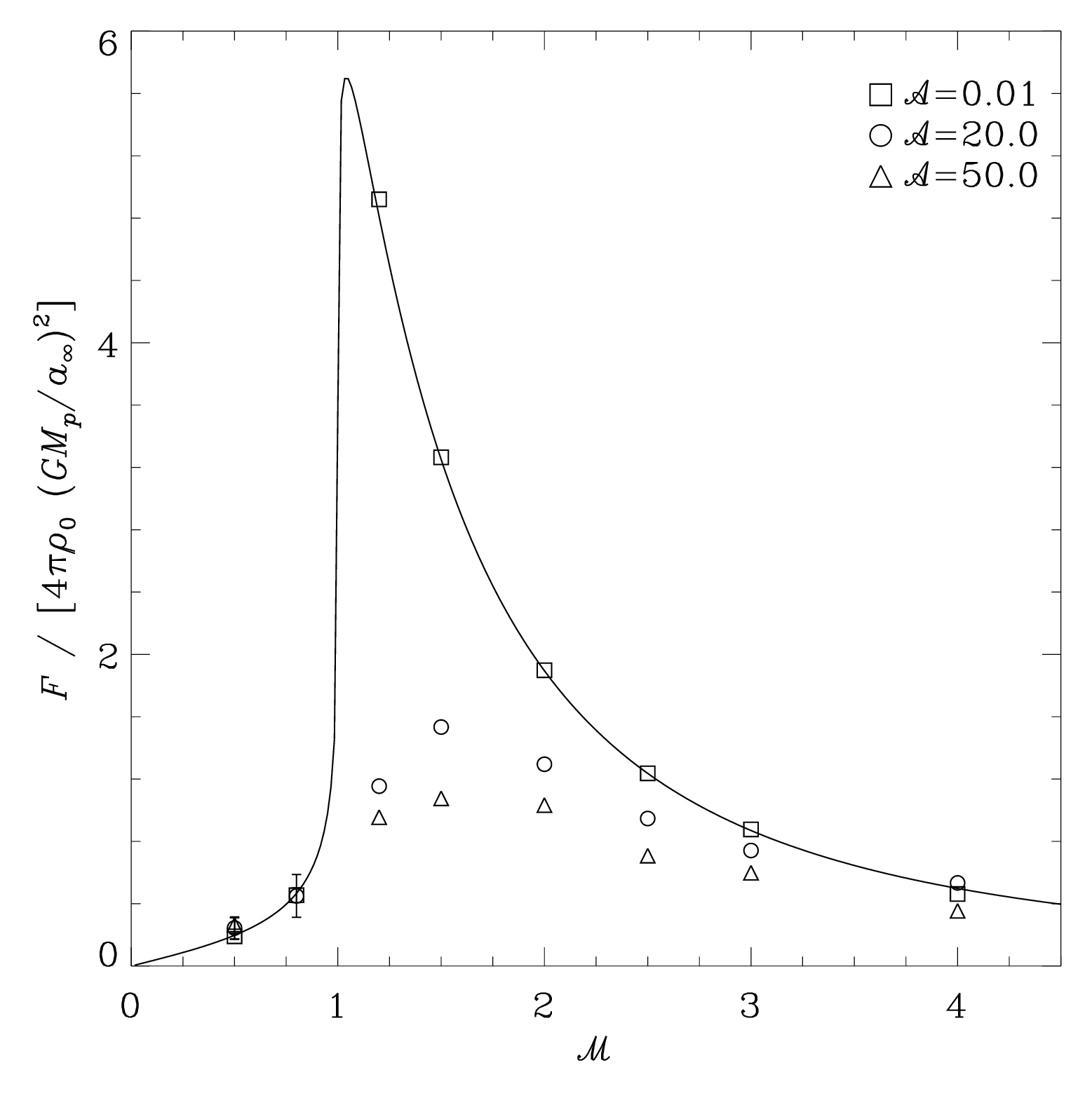}
  \caption{\label{fig:fM}
  Dimensionless DF forces for various models with different 
  $\A$ and $\mach$ at $t/\tcr=600$. 
  The solid line draws the linear drag force 
  (\eqo{eq:f_pt_linear}) with $\rmin=0.35\mach^{0.6}\rs$. 
  For supersonic models, 
  the drag force decreases with increasing $\A$. 
  Subsonic models do not reach a quasi-steady state and show
  some fluctuations, but their mean drag forces 
  are similar to the linear estimates with the same $\mach$.}
\end{figure}

\begin{figure}
  \epsscale{1}
  \plotone{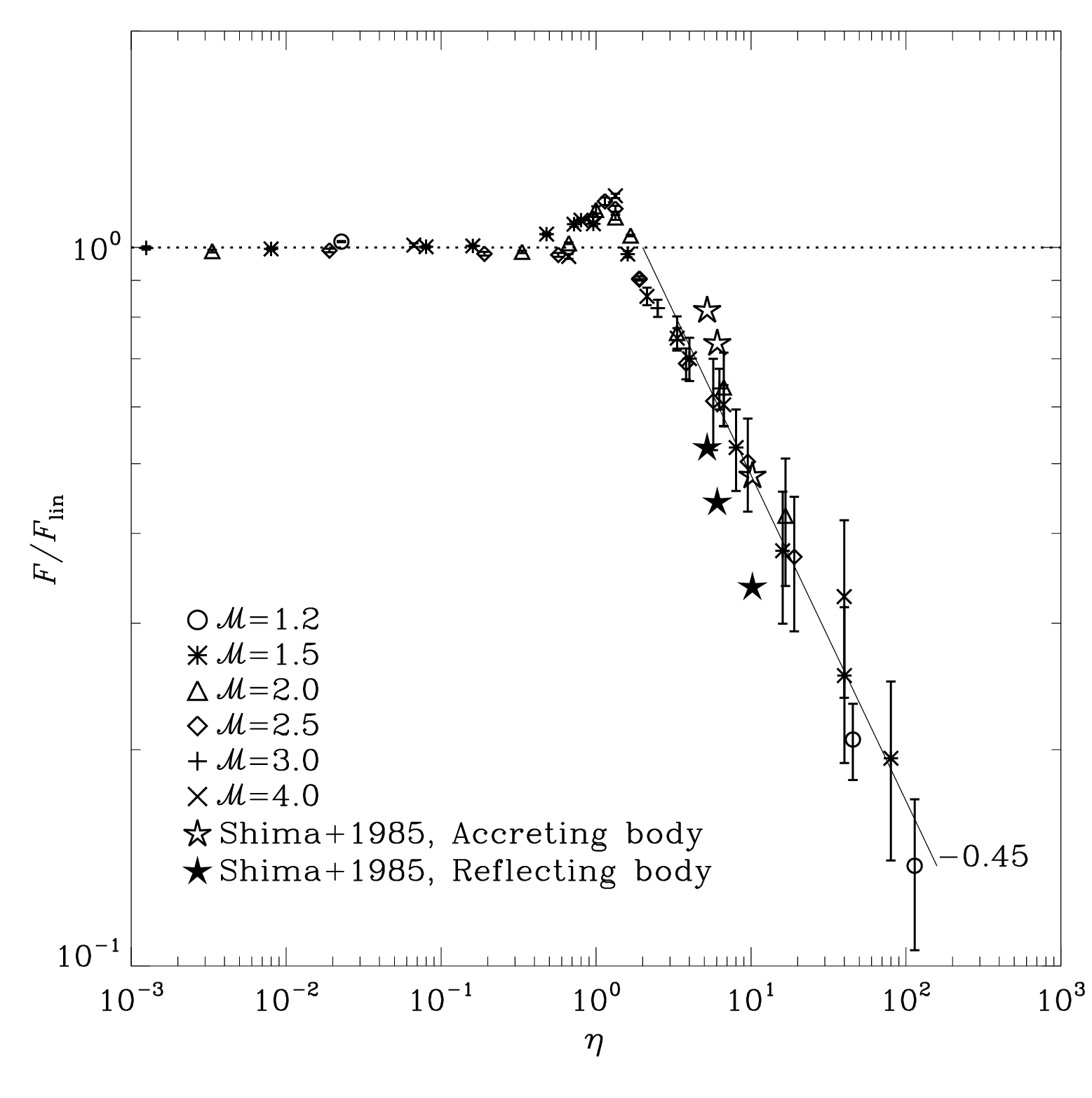}
  \caption{\label{fig:fA}
  Ratio of the nonlinear to linear DF forces for supersonic models
  as a function of the nonlinearity parameter $\Amod=\A/(\mach^2-1)$. 
  Various symbols and their errorbars indicate the means and 
  standard deviations of $F/\Fplin$ over time for $t/\tcr>50$.
  When $\Amod \la 0.7$, $F/\Fplin\approx 1$, while 
  $F/\Fplin\approx (\eta/2)^{-0.45}$ for $\Amod>2$.
  Also plotted as star symbols are the results of \citet{shi85} 
  for BHL flows onto accreting or reflecting bodies.}
\end{figure}

\begin{figure}
  \epsscale{1}
  \plotone{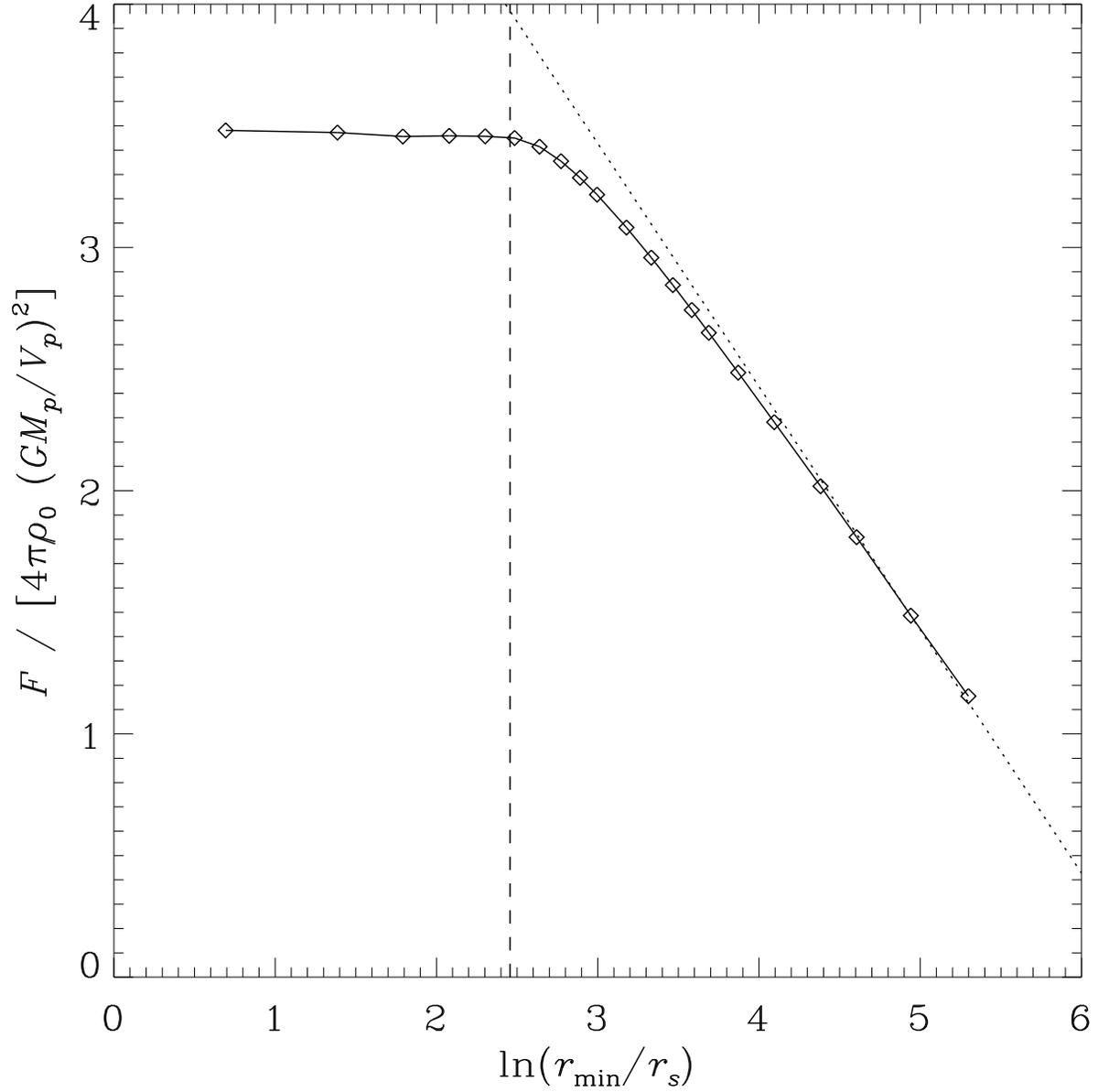}
  \caption{\label{fig:frmin}
  Dimensionless DF force as a function of the cut-off radius $\rmin$, 
  interior of which is excluded in the force evaluation, 
  for a model with $\A=20$ and $\mach=1.5$ at $t/\tcr=600$. 
  The vertical dashed line marks the detached shock distance of
  $\delta=12\rs$, while the dotted line indicates a slope of $-1$.
  The drag force is nearly constant for $\rmin \la \delta$ and
  decreases almost logarithmically at large $\rmin$.} 
\end{figure}

\begin{figure}
  \epsscale{0.8}
  \plotone{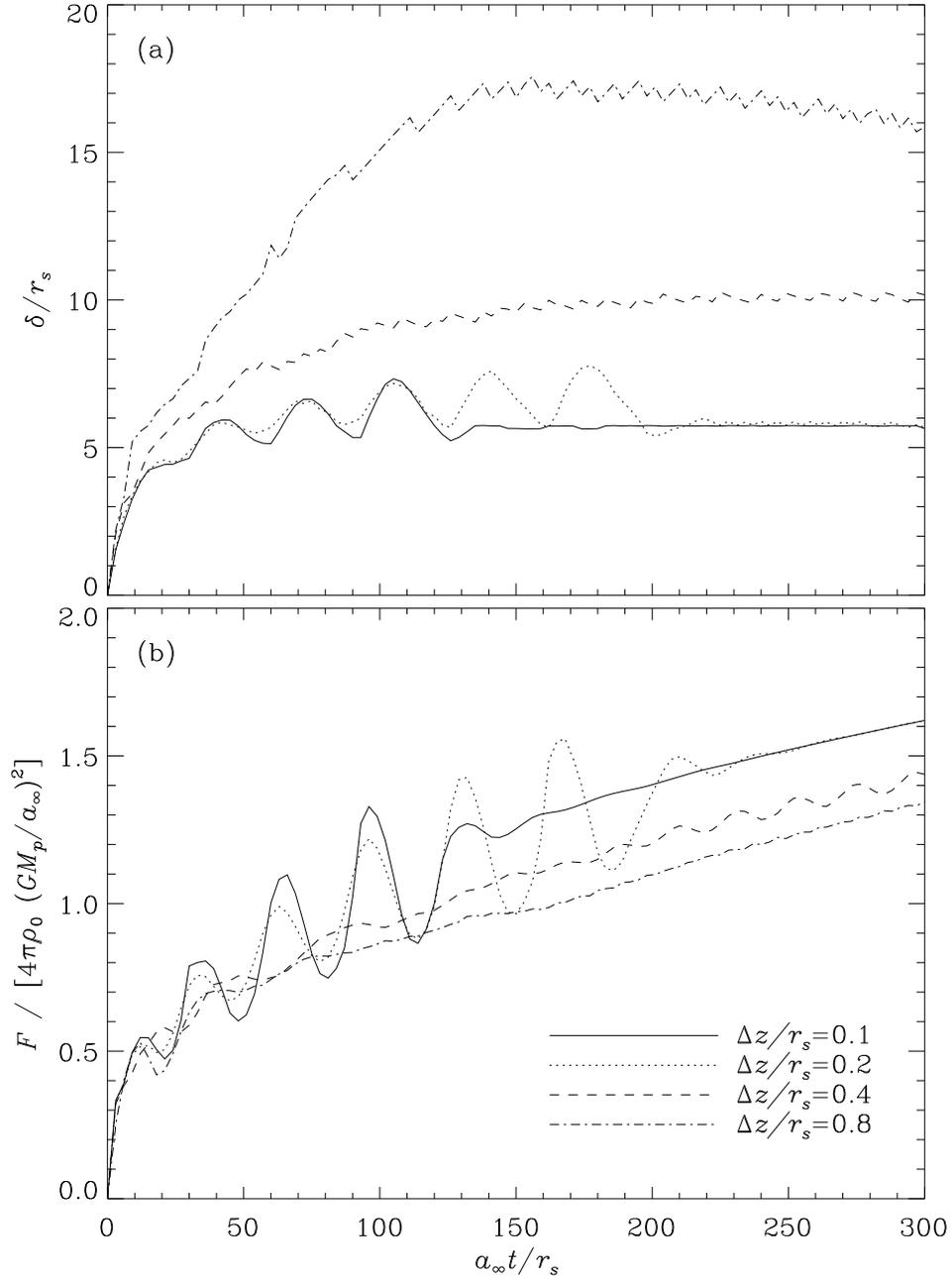}
  \caption{\label{fig:res}
  Time evolution of (\textit{a}) the detached shock distance $\delta$ and 
  (\textit{b}) the dimensionless DF force $F$ for $\A=10$ and $\mach=1.5$.
  The simulation results at four different resolutions are compared.
  Note that $\delta$ and $F$ are fully resolved if $\Delta z/\rs\ga 0.2$, 
  where $\Delta z$ is the grid spacing.}
\end{figure}

\begin{figure}
  \epsscale{1}
  \plotone{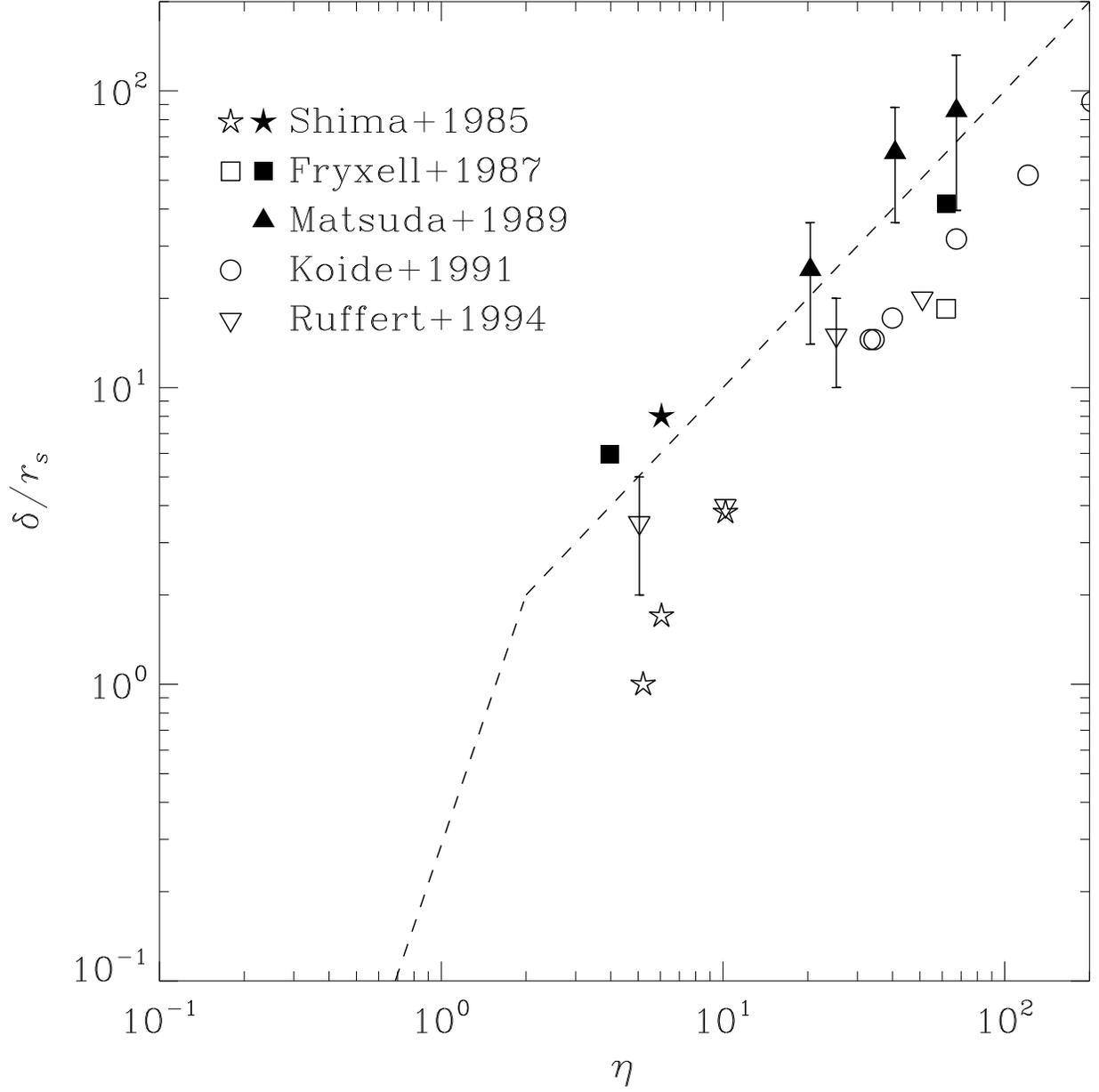}
  \caption{\label{fig:dEtaBHL}
  Detached shock distances against $\Amod$
  from the published work on the BHL accretion flows.
  Open symbols denote the results when the matter is allowed to be absorbed 
  to the perturber, while those under the reflection boundary conditions 
  are given by filled symbols.
  For comparison, the broken power laws that fit our numerical results 
  well are also shown as dashed lines.}
\end{figure}
\end{document}